\documentclass[10pt, conference, letterpaper]{IEEEtran}
\IEEEoverridecommandlockouts
\usepackage[T1]{fontenc}
\usepackage[utf8]{inputenc}
\usepackage{textcomp}
\usepackage{tabularx}

\usepackage{microtype}
\usepackage{amsmath,amssymb,mathtools}
\usepackage{graphicx} % For including images
\usepackage{booktabs} % For better table styling

\usepackage{algorithmic}
\usepackage{academicons}
\usepackage{graphicx}
\usepackage[caption=false,font=footnotesize,subrefformat=parens,labelformat=parens]{subfig}
\usepackage{xcolor}
\usepackage{verbatim}
\usepackage{pifont}
\usepackage{authblk}
\usepackage{siunitx}

\usepackage{xspace}
\usepackage{enumitem}
\usepackage{booktabs}
\usepackage{diagbox}
\usepackage[authormarkup=superscript,deletedmarkup=sout,addedmarkup=em]{changes}% here adding the option "final" would automatically generate the pdf without markup:
\usepackage{soul}
\soulregister\cite7
\soulregister\ref7
\soulregister\pageref7

\colorlet{soulcyan}{cyan!30}
\colorlet{soulgreen}{green!30}

\definechangesauthor[name={Claudio}, color=green!75!black]{CF}

\usepackage{xurl}

\usepackage{glossaries}
\newacronym{aoa}{AoA}{angle of arrival}
\newacronym{ap}{AP}{access point}
\newacronym{awv}{AWV}{antenna weight vector}
\newacronym{bpsk}{BPSK}{binary phase shift keying}
\newacronym{cots}{COTS}{commercial off-the-shelf}
\newacronym{csi}{CSI}{channel state information}
\newacronym{cnn}{CNN}{convolutional neural network}
\newacronym{dmg}{DMG}{directional multi-gigabit}
\newacronym{edmg}{EDMG}{enhanced directional multi-gigabit}
\newacronym{fl}{FL}{federated learning}
\newacronym{ftm}{FTM}{fine timing measurement}
\newacronym{fpbt}{BT}{beamforming training}
\newacronym{fpga}{FPGA}{field programmable gate array}
\newacronym{isac}{ISAC}{integrated sensing and communications}
\newacronym{iot}{IoT}{internet of things}
\newacronym{jcas}{JCAS}{joint communications and sensing}
\newacronym{los}{LOS}{line of sight}
\newacronym{mac}{MAC}{media access control}
\newacronym{mimo}{MIMO}{multiple-in multiple-out}
\newacronym{ml}{ML}{Machine Learning}
\newacronym{mmwave}{mmWave}{millimeter-wave}
\newacronym{nlos}{NLOS}{non line of sight}
\newacronym{ofdm}{OFDM}{orthogonal frequency-division multiplexing}
\newacronym{phy}{PHY}{physical layer}
\newacronym{psta}{P-STA}{passive station}
\newacronym{ppdu}{PPDU}{physical layer protocol data unit}
\newacronym{pdp}{PDP}{power delay profile}
\newacronym{qpsk}{QPSK}{quadrature phase shift keying}
\newacronym{rssi}{RSSI}{received signal strength indicator}
\newacronym{sgd}{SGD}{stochastic gradient descent}
\newacronym{slam}{SLAM}{simultaneous localization and mapping}
\newacronym{soa}{SotA}{State of the Art}
\newacronym{sta}{STA}{station}
\newacronym{snr}{SNR}{signal noise ratio}
\newacronym{sdr}{SDR}{software defined radio}
\newacronym{tof}{ToF}{time of flight}
\newacronym{trn}{TRN}{training}
\newacronym{usrp}{USRP}{universal software radio peripheral}
\newacronym{ura}{URA}{uniform rectangular array}
\newacronym{waveslam}{waveSLAM}{mmWave-based simultaneous localization and mapping}
\newacronym{wlan}{WLANs}{wireless local area networks}

\newcommand{\nref}[2]{\ref{#1}\,\subref{#2}}

\newcommand{\ourmodel}{\textsc{MilaGro}}

\makeatletter
\newcommand\remembertext[2]{% #1 is a key, #2 is the text
  \immediate\write\@auxout{\unexpanded{\global\long\@namedef{mytext@#1}{#2}}}%
  {\color{blue} #2}% This is the *only* place it will be printed by this command
}
\newcommand\recalltext[1]{%
  \ifcsname mytext@#1\endcsname
    \@nameuse{mytext@#1}%
  \else
    ``???''
  \fi
}
\makeatother
\begin{document}

\title{802.11bf Multiband Passive Sensing: Reusing Wi-Fi Signaling for Sensing}

\author{Pablo Picazo-Martinez\textsuperscript{*}, 
    Carlos Barroso-Fern\'andez\textsuperscript{*}, 
    Alejandro Calvillo-Fernandez\textsuperscript{*}, \\
    Milan Groshev\textsuperscript{†}, 
    Carlos J. Bernardos\textsuperscript{*}, 
    Antonio de la Oliva\textsuperscript{*}, 
    Alain Mourad\textsuperscript{‡}}

\maketitle

\renewcommand{\thefootnote}{\fnsymbol{footnote}}

\footnotetext[1]{Pablo Picazo-Martinez, Carlos Barroso-Fern\'andez, Alejandro Calvillo-Fernandez, Carlos J. Bernardos, and Antonio de la Oliva are with the Department of Telematics of Universidad Carlos III de Madrid, Spain (email:papicazo@pa.uc3m.es;cbarroso@pa.uc3m.es;acalvill@pa.uc3m.es;
cjbc@it.uc3m.es;aoliva@it.uc3m.es)}
\footnotetext[2]{Milan Groshev is with IE School of Science and Technology, Spain (email:milan.groshev@ie.edu)}
\footnotetext[3]{Alain Mourad is Senior Director in InterDigital Europe, UK (email: alain.mourad@interdigital.com)}

\renewcommand{\thefootnote}{\arabic{footnote}}
\setcounter{footnote}{0}

% Optional acknowledgment or funding can go here:
\footnotetext{This work has been partially funded by the European Commission Horizon Europe SNS JU MultiX (GA 101192521) Project, the project 6GINSPIRE PID2022-137329OB-C42, funded by MCIN/ AEI/10.13039/501100011033/”, and the Regional Government of Comunidad de Madrid under grant agreement No. TEC-2024COM-360 (Disco6G).
THIS WORK HAS BEEN SUBMITTED TO IOTJ ©2025 IEEE. Personal use of this material is permitted. Permission from IEEE must be obtained for all other uses, in any current or future media, including reprinting/republishing this material for advertising or promotional purposes, creating new or redistribution to servers or lists, or reuse of any copyrighted component of this work in other work.}

%Accurate Indoor Mapping Through mmWave Joint Communications and Sensing with Off-the-Shelf Devices

%\author{Anonymous Authors}

 %\author{\IEEEauthorblockN{1\textsuperscript{st} Pablo Picazo-Martínez}
 %\textit{Universidad Carlos III de Madrid}\\
 %Madrid, Spain \\
 %\and
 %\IEEEauthorblockN{2\textsuperscript{nd} Milan Groshev}
 %\textit{Universidad Carlos III de Madrid}\\
 %Madrid, Spain \\
 %\and
 %\IEEEauthorblockN{3\textsuperscript{rd} Alejandro Blanco}
 %Madrid, Spain\\
 %\and
 %\IEEEauthorblockN{4\textsuperscript{th} Claudio Fiandrino}
 %\textit{IMDEA Networks Institute}\\
 %Madrid, Spain \\
 %\and
 %\IEEEauthorblockN{5\textsuperscript{th} Antonio de la Oliva}
 %\IEEEauthorblockA{\textit{Universidad Carlos III de Madrid} \\
 %Madrid, Spain\\
 %\and
 %\IEEEauthorblockN{6\textsuperscript{th} Joerg Widmer}
% \textit{IMDEA Networks Institute}\\
 %Madrid, Spain \\

\begin{abstract}
This paper presents a novel multiband passive sensing system that leverages existing Wi-Fi signals for environmental sensing, following IEEE 802.11bf standardization efforts. By combining Channel State Information (CSI) from sub 7GHz and mmWave, the system enhances single band passive sensing accuracy and reliability in detecting human presence, movement, and activities in indoor environments. Utilizing a novel model, called \ourmodel{}, the system demonstrates robust performance across different scenarios, including monitoring human presence in workspaces and tracking movement in corridors. Experimental results show how we  improved the performance of single band passive sensing approaches by integrating multiband data. The system also addresses key security concerns associated with passive sensing, proposing measures to mitigate potential risks. This work advances the use of Wi-Fi for passive sensing by reducing reliance on active sensing infrastructure and extending the capabilities of low-cost, non-intrusive environmental monitoring.

%In this paper, we show how to enable robust and accurate indoor Simultaneous Localization and Mapping (SLAM) with off-the-shelf mmWave communication devices. mmWave technology has proven excellent for SLAM in situations with visual obstruction like smoke, glass/mirror walls or insufficient light that are an impediment proper operation of traditional optical sensors. However, past research has utilized mmWave radar technology. By contrast, we propose \toolname, a solution that exploits current mmWave communication technology for indoor mapping in the spirit of Joint Communications and Sensing. Unlike radar systems that are built in purpose for detection and tracking, \toolname uses commercial-off-the-shelf communications devices for sensing. We see such an opportunity as modern robots may need to transmit high data rates sensory information which mmWave can afford. Our prototype implementation of \toolname and extensive field tests show that: \textit{i)} re-using communication technology for sensing at zero extra cost is feasible which allows avoiding the need for costly dedicated hardware crucial for micro-robots, \textit{ii)} the localization accuracy obtained is remarkably high (cm-level precision) for diverse wall materials.
\end{abstract}


\section{Introduction}
\label{sec:intro}

As of 2025, there are approximately 21.1 billion Wi-Fi-enabled devices in use worldwide, spanning various environments, including homes, enterprises, and industrial settings~\cite{WiFi_2024status}. This existing wireless infrastructure offers a unique opportunity to leverage the existing Wi-Fi signals in order to detect and interpret physical changes in the environment. The nature of the Wi-Fi signals is affected by objects and movements in a space, which can be measured to gain insights about user motion, occupancy, and even biometric information (like breathing or heart rate). As such, Wi-Fi sensing finds applicability in intrusion detection systems, remote health monitoring, and home automation.  

\begin{figure}
\centering
\includegraphics[width=\columnwidth]{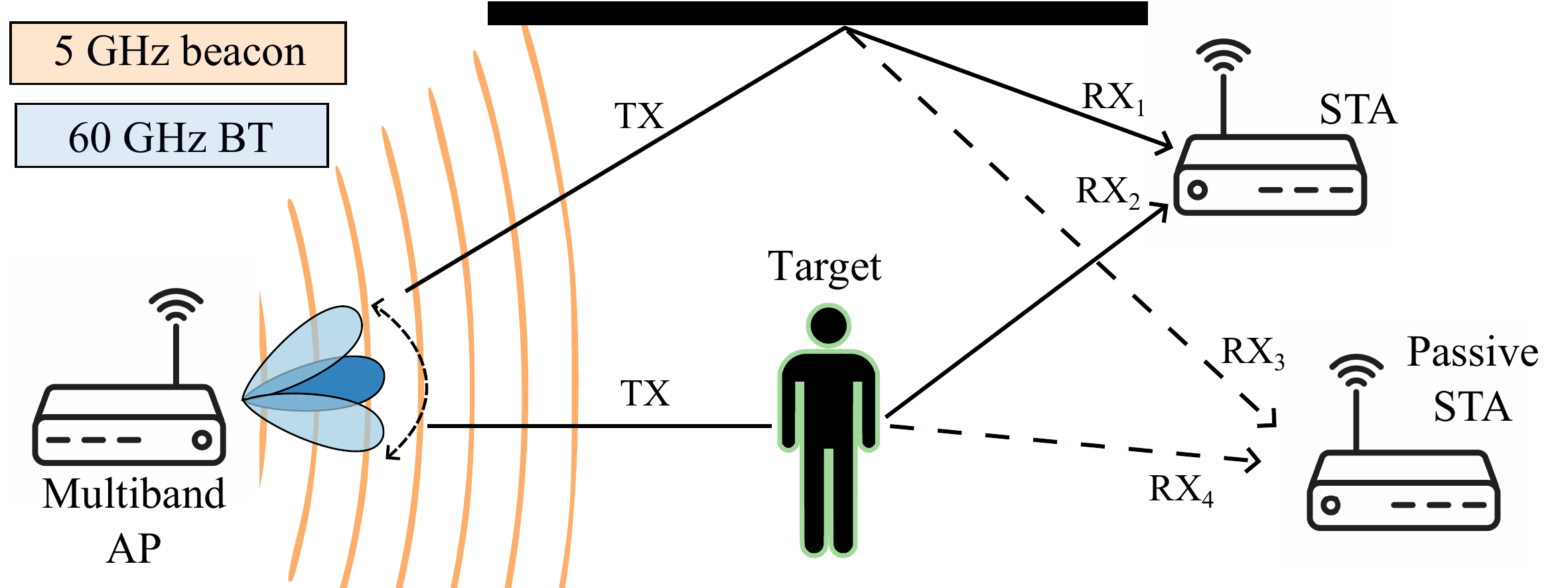}
\vspace*{0ex}
\caption{Multiband AP sends 5~GHz (orange) and 60~GHz (blue). STA uses these signals for communications (solid line) while Passive STA reuses them for sensing purposes (dashed line).}
\label{fig:intro}
\vspace*{-3ex}
\end{figure}

%standardization bodies who do sesing and different types of sensing and the motivation behind using passive sensing
Motivated by these opportunities the IEEE 802.11bf amendment introduces the needed modifications to the 802.11 standard to support Wi-Fi sensing. 
The IEEE 802.11bf amendment defines two main types of sensing configurations depending on the type of signal used:
\emph{1)} Active sensing uses specific signals to perform a sensing task; while this solution achieves higher accuracies, it consumes valuable resources such as spectrum and energy. \emph{2)} Passive sensing, uses existing Wi-Fi signals to perform a task; while this does not interfere with existing communications in the network, not requiring users in the network to perform any task, it has less capabilities than active sensing.

In this paper we focus on passive sensing. Particularly, we use signals such as 5 GHz beacons or mmWave \gls{fpbt}, which are valuable for sensing purposes -- see Fig.~\ref{fig:intro}, where a \gls{psta} takes advantage of communications between an \gls{ap} and the target \gls{sta} to infer the environment. These signals can be used either as a standalone solution or as a complement to other sensing technologies, enhancing overall accuracy and robustness.

%SoA works that do sub-6 passive sensing with examples and mmwave passive sensing with examples , each one with its problems

Recent works have applied passive sensing in both sub-7~GHz Wi-Fi and mmWave bands. In the sub-7~GHz range, \cite{passub61} and \cite{passub62} offer low-cost solutions but struggle with accuracy and sensitivity to environmental changes, even when advanced techniques like deep learning are used. Meanwhile, mmWave approaches, such as \cite{pasmmwave1} and \cite{pasmmwave2}, achieve higher accuracy but face drawbacks like long sensing times and complex hardware setups, limiting their practicality in real-time scenarios. These challenges highlight the need for a multiband approach to overcome the individual limitations of each technology.

%in this paper what we do and description of our system

\begin{table*}[t]
\centering
\label{tab:usecase}

\caption{IEEE 802.11bf Use Cases and Feasibility with Passive Sensing}
\begin{tabular}{|p{2.2cm}|p{3.5cm}|p{3.7cm}|p{3.2cm}|p{1.3cm}|}
\hline
\textbf{Use Case} & \textbf{Description of the task} & \textbf{Accuracy / KPIs needed} & \textbf{Periodicity / Refresh Rate} & \textbf{Passive?} \\
\hline
Presence Detection & Detect human presence or motion using CSI variations. & High reliability indoors; range $\sim$10--15 m. & $\sim$10--100 ms (depends on beacon/CSI rate). & Yes \\
\hline
People Counting / Room Sensing & Estimate number of people in a room by analyzing Wi-Fi channel changes. & Moderate accuracy ($\pm$1--2 people); sensitive to density. & 100 ms -- 1 s. & Yes \\
\hline
Healthcare Monitoring & Track breathing, heart rate, or falls through small CSI variations. & Fall detection feasible; vital signs accuracy low ($<$80\%). & Requires $<$50 ms for respiration/heartbeat, hard to achieve passively. & Insufficient \\
\hline
Gesture Recognition & Recognize hand/arm gestures or control appliances. & Large gestures feasible; fine gestures low accuracy ($<$70\%). & 5--20 ms for responsiveness. Passive sensing too slow for fine gestures. & Partially \\
\hline
Human Activity Recognition & Identify activities like walking, sitting, or fine movements. & Moderate (70--85\%), depends on multipath richness. & 50--200 ms. & Yes \\
\hline
In-Car Sensing & Detect occupants, monitor driver drowsiness, or children/pets. & Presence detection feasible; respiration/drowsiness limited. & 20--50 ms ideally, passive Wi-Fi often slower. & Insufficient \\
\hline
Localization / Object Tracking & Estimate range, velocity, position of people/objects with CSI. & Positioning error $\sim$1--2 m indoors. & 10--100 ms; passive less consistent than active. & Insufficient\\
\hline
\end{tabular}
\end{table*}

This paper proposes a multiband passive sensing approach combining information from sub-7~GHz Wi-Fi beacons and \gls{mmwave} beamforming training procedures. By fusing data from these two bands, we improve sensing accuracy and gain a more complete understanding of the wireless environment. Additionally, we introduce a novel \gls{ml} model architecture, called \ourmodel{}, designed to process the multiband data efficiently and address the limitations of single-band sensing methods.

%What kind of experimental analysis we do and what we found
We conducted multiple experiments to evaluate the performance of our multiband passive sensing system. First, we assessed its capabilities for both sub-7~GHz and mmWave bands independently, as well as the training performance of our novel model for multiband data fusing, \ourmodel{}. Next, we tested a lab scenario with the Multiband \gls{ap} placed both inside and outside a room to analyze how well the system detects changes in different environments. Lastly, we performed movement tracking in a corridor using existing 5~GHz \gls{ap} infrastructure, adding a mmWave \gls{ap} on top of it to emulate the behavior of a Multiband \gls{ap}.

The results demonstrated that the combination of sub-7~GHz and mmWave bands enhanced the system's accuracy, particularly in challenging scenarios where each band alone faced limitations in capturing detailed information. This multiband approach proved especially effective in environments with complex signal conditions, providing a more reliable sensing performance.

%the paper si organized as follows

The paper is organized as follows. Sec.~\ref{sec:background} explains different Wi-Fi procedures that can be used in passive sensing, Sec~\ref{sec:motivation} motivates the contributions of the paper. Sec.~\ref{sec:model} presents our novel model for multi-band sensing and Sec.~\ref{sec:proto} elaborates on the system and prototype implementation.  Sec.~\ref{sec:results} contains the results of the experimental evaluation. Sec.~\ref{sec:sec} elaborates on key security aspects to be considered. Finally, Section~\ref{sec:soa} presents the state of the art and positions our solution within it, while Section~\ref{sec:conclusions} concludes the article.

\section{Passive sensing}
\label{sec:background}

\begin{figure*}[ht]
\centering
\includegraphics[width=\textwidth]{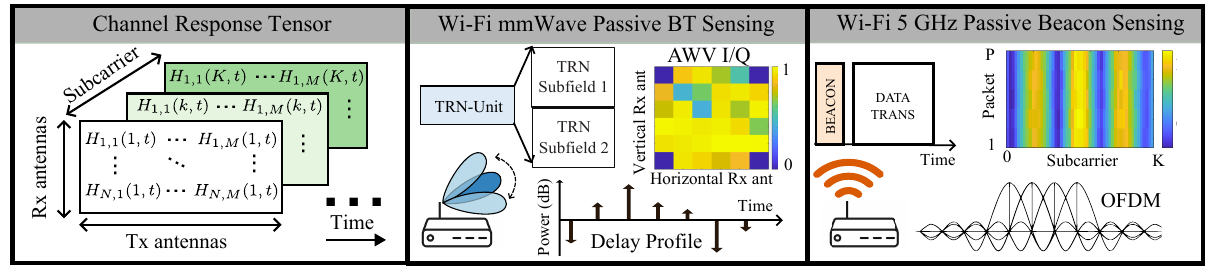}
\caption{ (Left) Channel Response for $N$ reception antennae, $M$ transmitting antennae and $K$ subcarriers at time $t$; tensor generalization. (Center) mmWave \gls{ap} performing \acrshort{fpbt}, transmitting the TRN-units; two graphs representing the I/Q for each receiving antenna, and the power of receiving signal over time, i.e., the \gls{pdp}, for an \acrshort{awv} configuration. (Right) and 5~GHz Wi-Fi \gls{ap} sending beacon frames and their respective I/Q at the \gls{psta}.}
\label{fig:s4}
\vspace*{-3ex}
\end{figure*}

%Wi-Fi sensing leverages existing wireless infrastructure to detect changes in the environment, such as the presence of objects or movement of people and animals, by analyzing variations in wireless signals. This capability transforms Wi-Fi from a communication tool into a versatile sensing technology, unlocking a wide range of applications across various industries.

In this section, we first introduce the use cases and KPIs proposed by standardization organizations \cite{3gpp,etsi,bf}, as well as the role of passive sensing within them. Afterwards, we explain the theoretical foundation of passive Wi-Fi sensing, followed by the procedures that can be employed for passive sensing in different frequency bands. In particular, we focus on sub-7~GHz beacons, and on \gls{fpbt} in \gls{mmwave} bands.

\subsection{Use cases enabled by passive sensing}

IEEE 802.11bf defines a variety of use cases that can benefit from this approach. These use cases span smart homes, healthcare, in-car monitoring, industrial environments, and localization applications. Table \ref{tab:usecase} shows the proposed IEEE 802.11bf use cases, and highlights their expected performance in terms of accuracy and refresh rate.  

Passive sensing can be adopted either as a complementary solution to existing sensing technologies or as a stand-alone approach, depending on the requirements of each use case. For example, in scenarios where high accuracy and robustness are critical, passive sensing may be combined with other sensing modalities to enhance reliability. On the other hand, in use cases with moderate accuracy needs and where cost-effectiveness or non-intrusiveness are prioritized, passive Wi-Fi sensing is a feasible solution to operate stand-alone. 

In this paper, we test whether our solution can achieve the expected performance as a standalone sensing tool in the following use cases: presence detection, people counting/room sensing, and gesture and human activity recognition.  

The goal is to provide a clear overview of the potential and limitations of passive Wi-Fi sensing in practical deployments.

\subsection{Theoretical foundations under Wi-Fi sensing}

As explained in~\cite{Cao2021}, let $X(f,t)$ and $Y(f,t)$ be the frequency domain representation of the transmitted and received signals with carrier frequency $f$ at time $t$, then
\begin{equation}
    Y(f,t) = H(f,t)X(f,t) + A(f,t),
\end{equation}
where $A(f,t)$ is the additive white Gaussian noise and $H(f,t)$ is the channel frequency response of the carrier. Notice that $H(f,t)$ is constructed with all paths of the signal. Hence, modifying the paths -- for example, introducing a new element in the environment -- will result in a disturbance of $H(f,t)$, and in consequence, in the received signal $Y(f,t)$.

Notice that the Wi-Fi \gls{csi} is precisely $H(f,t)$, defined as follows~\cite{WiFi-Sleep}:
\begin{equation}
    H(f,t) = \sum_{i=1}^L A_i e^{j2\pi \frac{d_i(t)}{\lambda}},
\end{equation}
where $L$ is the number of paths, $A_i$ the amplitude attenuation, $\lambda$ the wavelength, and $d_i(t)$ the length of the $i$ path at time $t$ (paths may change over time). In the case of multiple antennae, each antenna at reception $i$ receives a different signal from each transmitting antenna $j$, and perceives a slightly different channel response $H_{i,j}(f,t)$.
The \gls{csi} of the impulse response is typically expressed with tensors, considering the signal collected over time by each antenna and subcarrier, see Fig.~\ref{fig:s4}\,(left).

The degradation of the power density or path loss attenuation of a transmitted signal can be significantly influenced by various propagation phenomena such as free space loss, reflection or refraction. Several models in the literature manage to quantify such power density or path loss, each accounting for specific environmental conditions and propagation mechanisms.~\cite{PLmodel_sub6,PLmodel_mmWave,PL_mmWave}. 
%The reduction in power density of the transmitted signal -- or path loss attenuation (PL) -- is sensitive to different effects, as free space loss, reflection, refraction, etc. There exists different models to calculate it~\cite{PLmodel_sub6,PLmodel_mmWave,PL_mmWave}.
%\begin{equation*}
%    \text{PL} = P_{\text{TX}} - P_{\text{RX}},
%\end{equation*}
%where $P_{\text{TX}}$ and $P_{\text{RX}}$ are the transmitted and received power.
%\[
%x[n] = \frac{1}{N} \sum_{k=0}^{N-1} X_k e^{j2\pi kn/N}
%\]

As a result, we can affirm that the \gls{csi} contains valuable information about the environment and can be used for sensing~\cite{WiFi-Sleep}. However, extracting useful information from \gls{csi} is challenging unless the transmitted signal is known and controlled. In passive sensing, signals such as beacons or beamforming training do meet these requirements and are explained below.

\subsection{mmWave Beamforming Training}

IEEE 802.11ay contains enhancements to refine \gls{fpbt} procedures~\cite{beamtrain}. In \gls{mmwave} bands, \gls{sta}s need to know which is the best path to send the signal, as propagation characteristics of high frequencies do require accurate position estimations. To achieve this, in the \gls{fpbt}, the \gls{sta}s test various mmWave antenna configurations. Each configuration is specified by its \gls{awv}, which achieves a particular beam, using arrays that contain multiple antennas ($N$). As a result, the procedure sweeps from 60º to 120º depending on the antenna array used. Specifically, the \gls{sta}s scan multiple \gls{awv} configurations and choose the optimal one
\begin{equation}
    \text{AWV} = \left\{ (A_1,\phi_1),\ \dots,\ (A_N,\phi_N) \right\},
\end{equation}
being $A_n$ and $\phi_n$ the amplitude and phase of the signal transmitted by the $n$ antenna.

The packets transmitted contain a \gls{trn} field used for channel estimations during the \gls{fpbt} procedure. This \gls{trn} field includes \gls{trn}-Units each comprising \gls{trn} Subfields filled with Golay sequences. Each \gls{trn}-Unit corresponds to a particular \gls{awv}. At reception, these fields generate a \acrfull{pdp}, which contains the in-phase, quadrature (I/Q), and \gls{snr} of each tap. Analyzing these \glspl{pdp}, \gls{sta}s can derive which is the best \gls{awv}, and thus, the best path (angle and azimuth) between the \gls{sta}s involved (see Fig.~\ref{fig:s4}\,(center)).

A \gls{psta} can not derive the best path between the two \gls{sta}s by listening to this procedure, as it is encrypted. However, this information can be useful for sensing purposes, as we can know if a particular path between the \gls{sta} and the \gls{psta} is obstructed. Multiple paths are tested during the beam sweep procedure, providing valuable environmental information. As mmWave \gls{sta}s use \gls{fpbt} procedure periodically when transmitting data, to ensure that the beam points to the target \gls{sta}, \gls{psta} can collect this information without affecting communications, deriving useful information for environmental sensing.

\subsection{Signals for 5~GHz passive sensing}

In passive sensing, it is necessary to rely on constant and predictable signals to ensure accurate and consistent analysis. For this reason, we have chosen to focus on Wi-Fi beacon frames, which meet these requirements. Beacons are regularly transmitted by the \gls{ap} and serve an essential role in network management by broadcasting vital information to all devices within range. To maximize compatibility and reliability, they use either \gls{bpsk} or \gls{qpsk} as the modulation scheme. Additionally, beacons are always transmitted over a 20 MHz channel width across all Wi-Fi channels, ensuring that all devices, regardless of their specific Wi-Fi standard or capabilities, can successfully receive and decode them, as shown in Fig. \ref{fig:s4}\,(right).

This consistent use of 20 MHz for beacon frames makes them highly suitable for passive sensing applications, as the \gls{psta} can rely on the known, repeated transmission of information from the \gls{ap}. Although the resolution of beacon frames may be lower in comparison with other active sensing methods, their non-intrusive nature offers a cost-effective solution that does not interfere with communications, making them ideal for real-world sensing applications.

From this point forward, all references to sensing in this manuscript will refer to passive sensing.

\section{Motivation}
\label{sec:motivation}

%antonio liked this paragraph here mot

Tracking individuals in an office using Wi-Fi signals can provide significant advantages. By using the existing Wi-Fi infrastructure, organizations can gain valuable insights into employee movement patterns, identifying when and where individuals enter different areas within the office. Ultimately, Wi-Fi-based tracking offers a non-invasive and cost-effective solution for optimizing office environments while respecting privacy.

%sub6
Current passive sensing systems mainly use sub-7~GHz frequencies, especially 2.4, 5, and 6~GHz Wi-Fi bands. These frequencies support applications like human activity recognition and environmental monitoring, benefiting from Wi-Fi's widespread presence and ability to penetrate walls, which allows for larger indoor coverage areas without requiring \gls{los}. However, sub-7~GHz sensing has limitations that decrease its accuracy in applications that require high granularity. The lower spatial resolution due to longer wavelengths makes it harder to detect small movements or details, such as in gesture recognition. Additionally, limited bandwidth restricts detailed analysis, and the frequency bands are congested with interference from Bluetooth and IoT devices, impacting reliability in environments with many wireless signals.

%mmwave
mmWave frequencies, such as the 60~GHz band used in Wi-Fi, offer shorter wavelengths that improve spatial resolution and wider bandwidth (starting at 2.16~GHz) for capturing fine details, enabling the detection of small movements and gestures that sub-7~GHz systems miss. mmWave also faces less interference due to operating in less congested spectra, enhancing reliability even in dense wireless environments. However, mmWave has drawbacks, including limited range and poor penetration through materials like walls, restricting its sensing to closer objects and line-of-sight scenarios. Its narrow beam width further limits the field of view and covered area.

Next, we evaluate the sensing capabilities of each band to perform people detection and tracking. Based on this, we perform three experiments for both sub-7~GHz and mmWave: (i) people detection based on the \gls{psta} distance, (ii) people detection based on \gls{psta} orientation and (iii) people monitoring at best \gls{psta} location, derived previously. All these experiments where analyzed through a state-of-the-art \gls{cnn} with small adaptations for each use case\footnote{https://github.com/Bengal1/Simple-CNN-Guide}.

\subsection{Distance test}

\begin{figure}
    \subfloat[]{%
        \includegraphics[width=%
        .5\columnwidth]{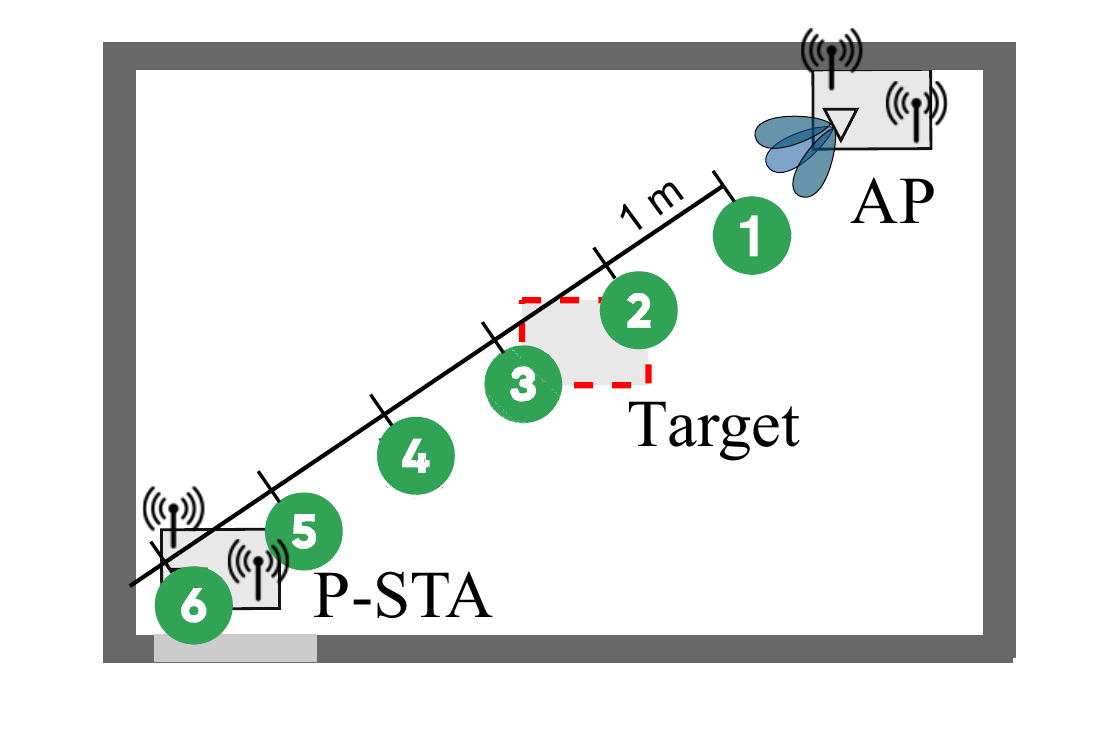}%
        \label{fig:scenarios2}%
    }%
    \subfloat[]{%
        \includegraphics[width=%
        .5\columnwidth]{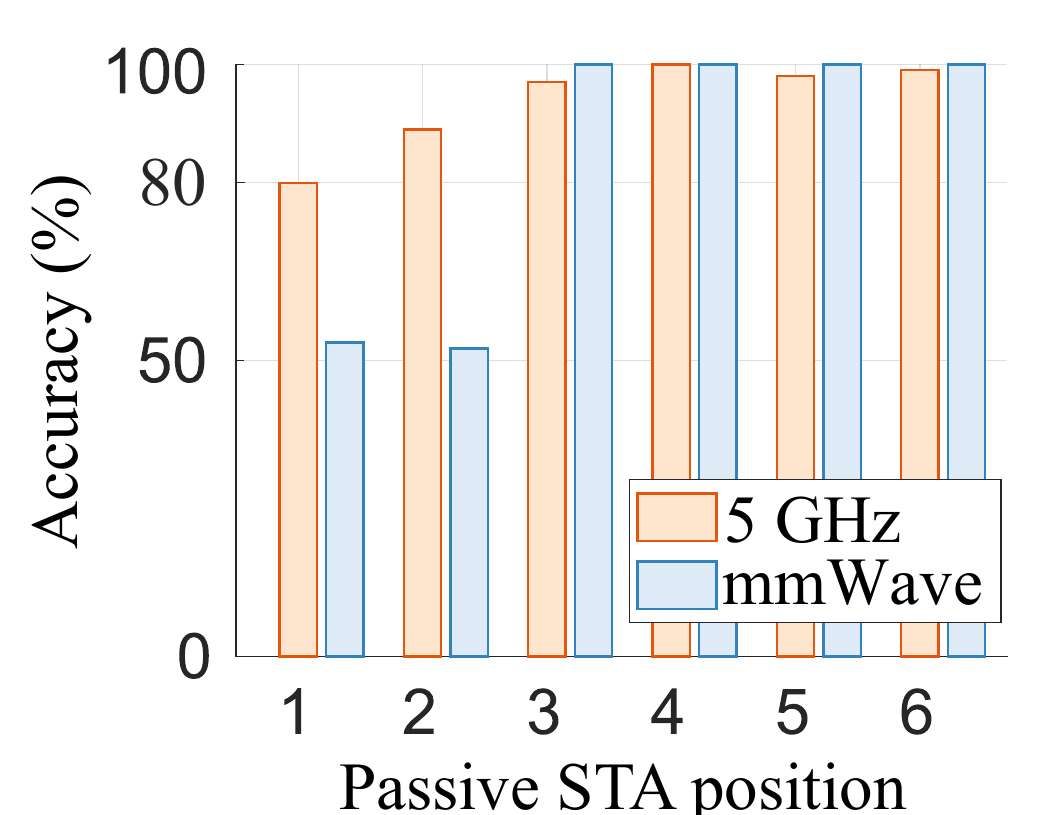}%
        \label{fig:distance}%
    }%
    \caption{Scenario and results for the distance test. Notice the \gls{psta} is between the target and the \gls{ap} in positions 1 and 2.}
    \label{fig:cap_distance}
\end{figure}

%setup, and data collection
We evaluate the accuracy of detecting a target centered in a room, and we place the \gls{psta} at different distances, one per testing point, from 1 m  to 6 m in a diagonal, and the detection target at two meters distance froom the \gls{ap}, located as shown in Fig.~\nref{fig:cap_distance}{fig:scenarios2}. Note that in some cases, the target is not between the \gls{ap} and the \gls{psta}. In this experiment, we consecutively collected 100 measurements of \gls{csi} and \gls{fpbt} for sub 6 GHz and mmWave, where the target is always in the same position.

%plot explanation + discussion and observations
To evaluate the system's performance, we define accuracy as the proportion of correct target detections within a given number of samples. For this experiment, we split 80\% of data on training and 20\% on test. Results show that if the \gls{psta} is behind the target the accuracy for 5~GHz is close to 100\%, and 100\% for mmWave. However, if the receiver is located between the target and the \gls{ap}, 5~GHz accuracy is reduced to 80\% and mmWave to 50\% (see Fig.~\nref{fig:cap_distance}{fig:distance}). This can be explained due to the propagation characteristics of mmWave, which is highly directional and does not work properly with \gls{nlos}.

\subsection{Orientation test}

\begin{figure}
    \subfloat[]{%
        \includegraphics[width=%
        .5\columnwidth]{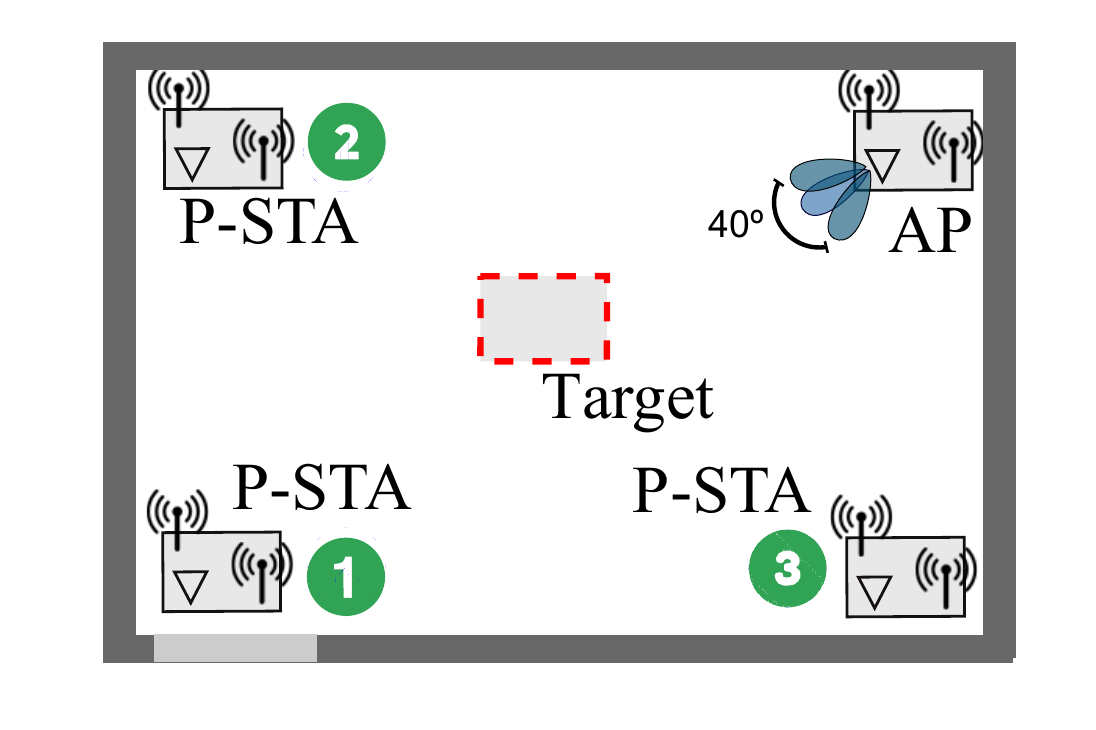}%
        \label{fig:scenarios1}%
    }%
    \subfloat[]{%
        \includegraphics[width=%
        .5\columnwidth]{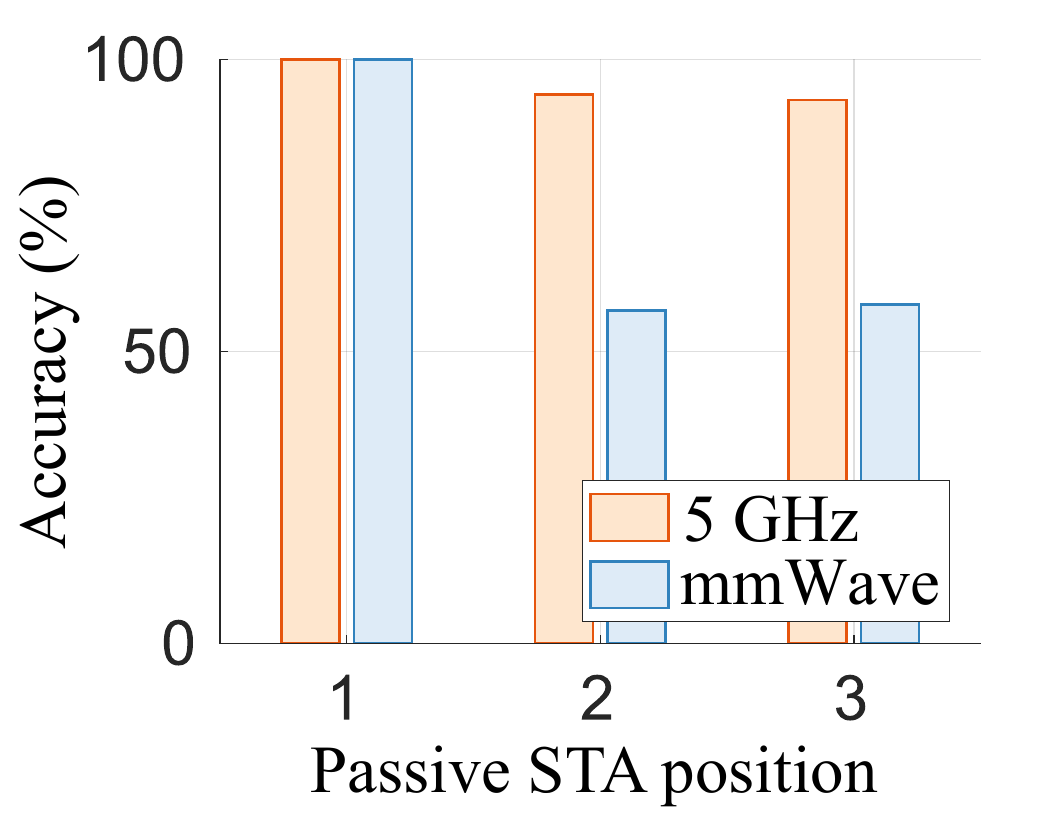}%
        \label{fig:orientation}%
    }%
    \caption{Scenario and results for the orientation test. The target do not interfere the \gls{los} path for \gls{psta} positions 2 and 3.}
    \label{fig:cap_orientation}
\end{figure}

%goal
Next, we conducted an analysis of human presence detection depending on the orientation of \gls{ap} and the \gls{psta}. The \gls{psta} is located now at three different positions: diagonal at 5m (pos 1), vertically at 3m (pos 2), and horizontally at 4m (pos 3). The same number of measurements and training split from the previous experiment is adopted in this one. Note that, in some cases, the \gls{psta} is out of the beamwidth of the mmWave \gls{ap}.
%setup, and data collection
We evaluate the accuracy of detecting a target centered in a room, depending on the position of the passive receiver. We place the \gls{ap} on the corner of the room and locate the receiver on the other three corners, as Fig.~\nref{fig:cap_orientation}{fig:scenarios1} shows. 

%plot explanation + discussion and observations
Results indicate that both 5~GHz and mmWave can individually detect the target properly with 100\% accuracy for position 1. In the case of positions 2 and 3, the accuracy at 5~GHz is reduced but stays above 90\%; this can be explained because the signal suffers from reflections when it reaches these positions. However, in the case of mmWave, the signal received is not strong enough to detect the target properly as the beam has \gls{nlos} and receives reflections that are harder to classify. As a result, the accuracy of mmWave for this setup is 52\%, see Fig. \nref{fig:cap_orientation}{fig:orientation}.

\subsection{People monitoring}

\begin{figure}[ht]
\centering
\includegraphics[width=\columnwidth]{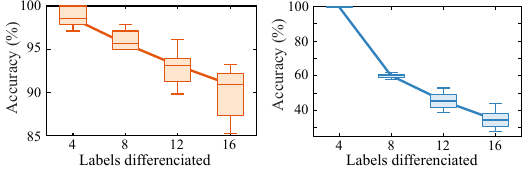}
\caption{(left) Sub-7~GHz accuracy against the number of labels to classify by the model and 
(right) mmWave \gls{fpbt} accuracy against the number of labels to classify by the model.}% Caption at the bottom
\label{fig:motivation}
\end{figure}

To evaluate the performance of sub-7~GHz and mmWave sensing we monitor human presence in a laboratory. In this scenario, there are four workstations, where people can be present or absent. This results in sixteen different people position combinations, each one represented by a label.
Following the previous experiments approach, we split 80/20\% for train/test and collect all the data consecutively.Accuracy is defined as the ratio of correctly predicted positions to the total number of positions. Based on this definition, we consider a misdetection to occur whenever the model outputs any combination different from the correct one—that is, when one or more specific positions are incorrectly classified as false positives or false negatives.
Fig.~\ref{fig:motivation} shows the accuracy obtained for both sub 7 GHz and mmWave using the state-of-the-art \gls{cnn}. For sub-7~GHz, the results show a decrease in accuracy as the number of classified labels increases. We can appreciate this effect in Fig.~\ref{fig:motivation}\,(left). As a result, we determined that further enhancements must be made for complex scenarios to increase the number of classified labels.

In \cite{surveybeam}, mmWave \gls{fpbt} is explored as an alternative to offer positioning. Motivated by this, we performed an analysis on how effective is  \gls{fpbt} to perform sensing. We replicated the previous experiment, classifying the 16 labels using mmWave information. 

Fig.~\ref{fig:motivation}\,(right) represents the accuracy obtained with mmWave \gls{fpbt}.

We observe that when the number of labels to classify is high, the accuracy is reduced. This happens because the beamsweep does not cover the whole tracked area, not detecting people who do not break \gls{los} between \gls{ap} and \gls{psta}.

\begin{figure}
\centering
\includegraphics[width=\columnwidth]{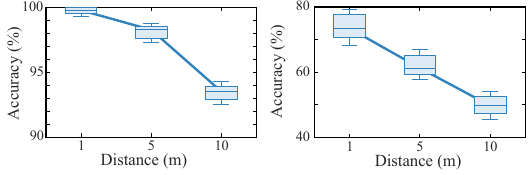}
\vspace*{-3ex}
\caption{mmWave accuracy detecting a single human target breaking LOS (left) and NLOS path (right) at 1,5,10 m.}
\label{fig:losnlos}
\vspace*{-3ex}
\end{figure}

To elaborate further on this, we decided to evaluate the performance of mmWave beamforming to detect humans breaking the \gls{los} or a \gls{nlos} path. Results in Fig.~\ref{fig:losnlos} show that as we increase the distance, \gls{nlos} cause big accuracy loss detecting a human presence. Thus, we validate that mmWave beamforming can not be used as a standalone tool for passive sensing.

The state of the art addresses these drawbacks by joining both bands for use cases such as positioning \cite{alejandro}, smartphone antennae \cite{smart}, or building multiband re-configurable intelligent surfaces \cite{ris}. However, none of these works consider using passive signals to extract information about the environment, offering a cost-effective solution without adding communication overhead.

Thus, we decided to design a system that combines information from multiple bands to exploit the inherent benefits of each band using a passive sensing approach. While sub-7~GHz offers better penetration, \gls{mmwave} \gls{fpbt} can assist with high-resolution measurements when the target breaks the \gls{los}. 

Our solution, presented in Section \ref{sec:model}, combines \gls{mmwave}’s precision with sub-7~GHz’s broader coverage through obstacles to provide a more comprehensive sensing system. %The next section will elaborate on the multiband passive sensing system design. 

\section{Multi-band Intelligence LeArning for Generalized Recognition and Observation (MilaGro)}
\label{sec:model}

%\section{Heterogeneous-band fusiOn for sensoRY tasks (HARMONY) ML model}

In this section, we present Multi-band Intelligence LeArning for Generalized Recognition and Observation (\ourmodel{}), a \gls{ml} model that uses the information from 5~GHz beacons and mmWave \gls{fpbt} to enable people detection and movement tracking. Due to the great performance and large validation of \acrfull{cnn} models in extracting valuable information from \gls{csi} data, \ourmodel{} integrates convolutional layers -- detailed below -- to achieve good accuracy in well-known state-of-the-art use cases, as presented in Sec.~\ref{sec:proto}.

\begin{figure}[ht]
\centering
\includegraphics[width=\columnwidth]{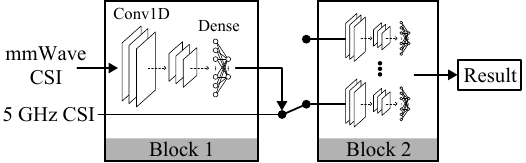}
\caption{Scheme of \ourmodel{}. mmWave data is introduced into a first block. The result is merged with the 5~GHz data to select the path to follow within the second block.} % Caption at the bottom
\label{fig:model}
\end{figure}

\gls{ml} models from the literature miss the integration of the information from different bands. Their approaches typically integrates \glspl{cnn} to extract information from one band.
\ourmodel{} highlights for processing data from 5~GHz and mmWave bands through its two building block structure (see Fig.~\ref{fig:model}). The first building block analyses the \gls{fpbt} data from the mmWave band to perform a coarse pre-classification, reducing the final inference space.
%Once this initial classification is made, 
This output is fed into the second building block, which incorporates the results of the first block with the information from the 5~GHz band. %This second block is composed by a set of convolutional layers, but with reduced complexity -- %as it leverages the output from the mmWave-based classification to refine its analysis.
%As previously remarked, mmWave data is used to perform a pre-classification that is refined with 5~GHz band data.
%This makes the process more efficient while maintaining high accuracy in interpreting the signals, achieving near real-time inference to enable time-sensitive use cases.
The result of the second block is the inference for the scenario which depends on the use case (e.g., the presence of people or the track of a person).

%Our design employs a two-stage classification process (see Fig.~\ref{fig:model}). First, we use a \gls{cnn} model to analyse the mmWave \gls{fpbt} data from the mmWave band to classify whether a path is obstructed or not. Once this initial classification is made, the result is fed into a second \gls{cnn} model, which incorporates the result of the first \gls{cnn} processing to the information from the 5~GHz band. This second \gls{cnn} model has reduced complexity, as it leverages the output from the mmWave-based classification to refine its analysis, making the process more efficient while maintaining high accuracy in interpreting the signals.

\subsection{Model architecture}
\label{subsec:modelarchitecture}
The layers of the two blocks of \ourmodel{} are summarized in Table~\ref{table:CNN}. The first block starts by applying a 1D convolutional layer (\textit{Conv1D}). The \textit{Conv1D} layer slides a filter along one dimension across the input sequence; the $i,j$ tensor from the output of the layer, named out$(i,j,:)$, is calculated by:
\begin{equation}
    \text{out}(i,j,:) = b(j) + \sum_{c=0}^{C-1} w(j,c,:) \star \text{in}(i,c,:),
    \label{eq:conv}
\end{equation}
where in$(i,c,:)$ is the $i,c$ tensor from the input of the layer, $b(j)$ is the $j^\text{th}$ bias, $C$ the number of channels, $w(j,c,:)$ the $j,c$ weight tensor, and $\star$ is the cross-correlation operator:
\begin{equation}
    (f\star g)(\tau) \coloneq \int^\infty_{-\infty} \overline{f(t)} g(t+\tau) dt,
\end{equation}
or $f$ and $g$ functions.
%Remark that inputs and outputs have three dimensions: 
This layer is designed to process one-dimensional (1D) sequential data where inputs vary over time, in order to detect local dependencies between adjacent time steps. The convolution operation applies 64 filters, each of them with a kernel size of two, which means each filter will examine pairs of consecutive time steps to extract short-term dependencies. By using 64 different filters, all learned features are represented as 64 distinct feature maps, each highlighting a different aspect of the input data. In other words, the first layer of the first block can learn a variety of features in parallel and how they affect the output.

%The layers of the building blocks of \ourmodel{} are summarized in Table~\ref{table:CNN}. It starts by applying a 1D convolutional layer (\textit{Conv1D}). 
%The \gls{cnn} is designed to process one-dimensional (1D) sequential data where inputs vary over time. 
%The \textit{Conv1D} layer slides a filter along one dimension across the input sequence, in other words, the $i,j$ tensor from the output of the layer, named out$(i,j,:)$, is calculated by:
%\begin{equation}
%    \text{out}(i,j,:) = b(j) + \sum_{c=0}^{C-1} w(j,c,:) \star \text{in}(i,c,:),
%    \label{eq:conv}
%\end{equation}
%where in$(i,c,:)$ is the $i,c$ tensor from the input of the layer, $b(j)$ is the $j^\text{th}$ bias, $C$ the number of channels, %$w(j,c,:)$ the $j,c$ weight tensor, and $\star$ is the cross-correlation operator:
%\begin{equation}
%    (f\star g)(\tau) \coloneq \int^\infty_{-\infty} \overline{f(t)} g(t+\tau) dt,
%\end{equation}
%for $f$ and $g$ functions. Remark that inputs and outputs have three dimensions: 
%The role of this layer is to detect local dependences between adjacent time steps. The convolution operation applies 64 filters, each of them with a kernel size of two, which means each filter will examine pairs of consecutive time steps to extract short-term dependencies.% By using 64 different filters, the CNN can learn a variety of features in parallel, {\color{red} REWRITE} meaning that these learned features are represented as 64 distinct feature maps, each highlighting a different aspect of the input data. 

\begin{table}[ht]
\centering
\caption{\ourmodel{} layers and properties.} % Caption on top
\label{table:CNN}

\begin{tabular}{rlc}
\toprule
& \textbf{Layers} & \textbf{Properties}   \\
\midrule

\rotatebox{90}{\hspace{-1em}Block 1} & \hspace{-1.1em}\textbf{
\begin{tabular}{l}
Conv1D \\
MaxPool1D \\
Dense \\
Droput \\
Dense
\end{tabular}} &
\hspace{-1.1em}\begin{tabular}{c}
filters=64, kernel size=2 \\
pool size=2   \\
size=128, activation=ReLU   \\
rate=0.5   \\
size=labels, activation=softmax   \\
\end{tabular}  \\

%\textbf{Conv1D} & filters=64, kernel size=2 \\
%& \textbf{MaxPool1D} & pool size=2   \\
%\textbf{Flatten} & -   \\
%& \textbf{Dense} & size=128, activation=ReLU   \\
%& \textbf{Dropout} & rate=0.5   \\
%& \textbf{Dense} & size=labels, activation=softmax   \\
\midrule

\rotatebox{90}{\hspace{-1em}Block 2} & \textbf{
\hspace{-1.1em}\begin{tabular}{l}
Conv1D \\
MaxPool1D \\
Dense \\
Droput \\
Dense
\end{tabular}} &
\hspace{-1.1em}\begin{tabular}{c}
filters=64, kernel size=2 \\
pool size=2   \\
size=128, activation=ReLU   \\
rate=0.5   \\
size=labels, activation=softmax   \\
\end{tabular}  \\

\midrule
& \textbf{Backward propagation} & optimizer=Adam, rate=0.1\%  \\
\bottomrule
\end{tabular}
\end{table}

% Space between the table and the figure
%\vspace{0.3cm}

% Figure

After the convolution, the block applies a \textit{MaxPooling1D} layer with a pool size of two. This layer reduces the dimensionality of the feature maps while preserving the most critical information. In this case, max pooling works by sliding a window of size two over each feature map and selecting the maximum value in each window. 
\begin{equation}
    \text{out}(i,j,l) = \max_{m=0,...,K-1} \text{in}(i,j,s\cdot l + m),
    \label{eq:maxpool}
\end{equation}
where $K$ is the kernel size and $s$ is the stride of the sliding window. \textit{MaxPooling1D} helps make the model invariant to small shifts in the input data, which is important in signal processing, where small variations in the input might occur due to interferences.

%Following the pooling, the feature maps are flattened into a one dimensional vector, transforming the array structure from the output samples -- notice that the first dimensions from the inputs and outputs of \eqref{eq:conv} and \eqref{eq:maxpool} represents the batch of samples processed together -- into a format that can be fed into fully connected layers.

The next layer is a \textit{Dense} layer with 128 neuron configured with ReLU activation to perform high-level feature abstraction. \textit{Dense} layers connect each neuron to every neuron in the previous layer, acting as a high-capacity layer that integrates information from all the feature maps. The ReLU activation function is used here to introduce non-linearity, enabling the network to learn complex, non-linear relationship in the data.

After that, a dropout is applied with a rate of 0.5, which means that randomly disables 50\% of the neurons in the dense layer during training to force the model to learn more robust features, preventing overfitting.

The final layer in the block is another \textit{dense} layer. This layer uses a softmax activation function, which converts raw output scores into probabilities.

As mentioned at the beginning of the section, this first block performs a pre-classification which reduces the complexity of the problem for the second block. This is implemented by using the output of the first block to select a path for the 5~GHz beacon \gls{csi} within the second block. Therefore, this second block contains several paths for the data, each of them formed by the same layers than those described in the first block.

The output of the second block is the inference result of the input data. It is a probability vector with the same dimension as the number of classes for classifying. Hence, the value of each dimension is the probability of belonging to that class, for a given sample.

\subsection{Model Training}

Model training is the phase in \gls{ml} where the model adjusts its parameters -- or weights -- to generalize from a subset of initial trial samples. Parameters are modified by solving the following optimization problem:

\begin{equation}
    \min_{\textbf{w}} L(F_\textbf{w}(X),y),
    \label{eq:ML_training}
\end{equation}

where \textbf{w} are the model parameters, $X$ is set of trial samples, $F_\textbf{w}(X)$ is the output when introducing the samples in the model with parameters \textbf{w}, $y$ are the expected values for all samples, and $L(F_\textbf{w}(X),y)$ is a loss function to compute the distance between the computed and expected value. The problem is highly influenced by the definition of the loss function, however there exist several optimizers that are proven to perform effectively in almost all scenarios~\cite{SGD_survey}, such as \gls{sgd}, which propose to update the parameters iteratively following:

\begin{equation}
    \textbf{w}^{k+1} = \textbf{w}^k - \eta \nabla L(F_{\textbf{w}^k}(X),y),
\end{equation}

being $\eta$ the size of each step, or learning rate, and $k$ representing the rounds, named epochs.

%{\color{red} REMOVE FL?}
In a distributed scenario, one can leverage different data sources to train the same \gls{ml} model, opening a bunch of new challenges. 
%\gls{fl} arise as a collaboratively way train the model without sharing raw data with others~\cite{FL_survey}, avoiding a security risk. It modifies objective function from problem \eqref{eq:ML_training} into
To address this new problem, the objective function from~\eqref{eq:ML_training} is modified into
\begin{equation}
    \hat{L}(\textbf{w}_1,\dots,\textbf{w}_N) = \frac{1}{N} \sum_{i=1}^N L_i(F_{\textbf{w}_i}(X_i),y_i),
\end{equation}
where $N$ is the number of collaborators, and sub-index $i$ indicates that the element belongs to $i^\text{th}$ collaborator. 
%In the original \gls{fl} scheme, the collaborators train its local models individually, and then they send the model parameters to a centralized server, which aggregates them through averaging and sends them back. 
Currently, in the literature, there exists articles that look to improve the process, tackling the main drawbacks: data privacy, centralization, performance under non-optimal conditions, synchronization, etc.~\cite{Survey_DL,FL_survey}.

In training mode, we back-propagate the error loss using a categorical cross-entropy function. The objective of this function is to measure the distance (or dissimilarity) between the predicted distribution and the real distribution of the classes. Then, based on the loss value from the previous function, we adjust the weights of the neural network using the \textit{Adam} optimizer~\cite{ADAM} with a learning rate of 0.001.

The following sections detail the system and prototype implementation designed to demonstrate the capabilities of the described model for multi-band passive sensing.

\section{MILAGRO testbed and prototype implementation}
\label{sec:proto}

This section presents our testbed and prototype to validate \ourmodel{} in real-world scenarios. We can differentiate three different parts: existing WLAN signals, Multiband \gls{psta}, and use cases enabled.

\begin{figure*}
\centering
\includegraphics[width=.9\textwidth,keepaspectratio]{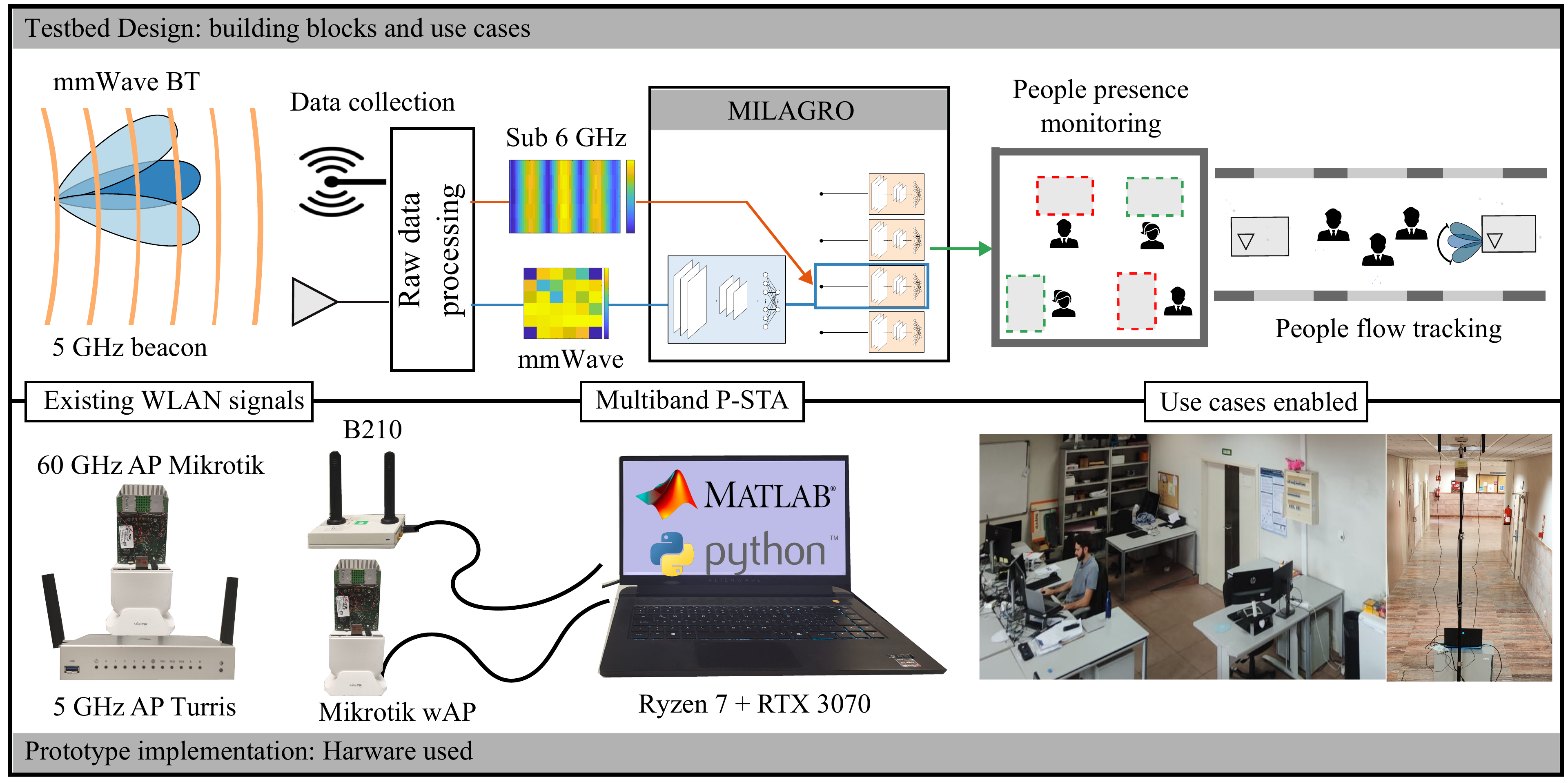}%
\vspace*{-1ex}%
\caption{Testbed building blocks (top) and hardware used for the implementation (bottom).}%
\label{fig:blocks}%
\vspace*{-3ex}%
\end{figure*}

\subsection{Existing WLAN signals generation} Our system relies on two bands: the 5~GHz band, where we sense 20~MHz \gls{ofdm} beacons, and the mmWave 60~GHz band, where we sense \gls{fpbt} procedures over 2.16~GHz bandwidth. Combining the broad coverage and obstacle penetration of sub-7~GHz with the high spatial resolution, increased sensitivity, and reduced interference of \gls{mmwave}, a multiband approach addresses the inherent shortcomings of current sub-7~GHz passive sensing systems. Having in mind that mmWave Wi-Fi devices are usually deployed together with sub-7~GHz networks to be able to fall back to the latter whenever the mmWave link is unavailable, we designed a testbed (see Fig. \ref{fig:blocks}) that simultaneously makes use of multiband signals to deliver a comprehensive environmental analysis.  

\textbf{5~GHz.}
To transmit the signal, we can either deploy a router, in our case a Turris Omnia 802.11ax, or listen to an existing router in the infrastructure. We configured our router to transmit in channel 48 of 5~GHz, but we can work in any band below 7~GHz. Despite our system allowing existing routers in the environment, regardless of the band used, we deploy this router to control variables such as location or data traffic when the experiment requires it. Despite beacons being control signals, they are transmitted using the same OFDM technology employed for data communication, with 20 MHz bandwidth for legacy compatibility.

\textbf{mmWave.}
To transmit the signal we use a commercial Mikrotik @60GHz router, equipped with a 6x6 \gls{ura}, this router is utilized to extract the result of the \gls{fpbt}, enabling detailed sensing over the 60GHz band, more details about the form of this procedure are available in \cite{az}. This router facilitates the obtention of necessary \gls{csi} components for our experimental purposes.

\subsection{Multiband P-STA}
These signals are processed in our Multiband P-STA, which consists on the following blocks: 

\textbf{Data collection}. This block captures sensing information from the 5~GHz and \gls{mmwave} bands. For the 5~GHz band, we collect Wi-Fi beacons, while for the \gls{mmwave} band, we listen to \gls{fpbt} frames. During this stage, we extract \gls{csi}  from the captured signals, which serve as the basis for further processing and analysis. 

It is important to highlight that \ourmodel{} operates with existing environmental signals, implying that the achievable accuracy depends on factors such as signal periodicity and signal density. Moreover, mmWave and 5 GHz signals do not require tight synchronization, since the fusion is performed in two independent layers, which tolerates deviations in signal collection of up to 100 ms in our implementation.

The \gls{psta} can be configured to capture signals with a defined periodicity. Each captured beacon packet contains 52 subcarriers, each with its corresponding complex values. Each \gls{fpbt} provides received signal strength per antenna (in our case with a rectangular 6x6 array) over a full sweep covering 60 degrees, (total size of about 90 kB). In our case, after preprocessing the files, the input for sub 6 GHz is 52 by 100 and for mmWave, 64 by 3000 (number of power measurements per \gls{awv} tested during the duration of a sweep).

\textbf{5~GHz.}
Although it is possible to collect the I/Q components using commercial routers, these implementations are not as accurate and efficient as using \gls{sdr} \cite{axcsi}. As a result, to collect the signal, we use the \gls{usrp} B210 \gls{sdr} combined with MATLAB libraries, as it is an efficient solution for data processing that allows real-time collection of I/Q samples before being decoded. These beacons provide crucial information about Wi-Fi activity and allow us to perform detailed signal analysis.

The USRP receiver captures the waveform, where beacons are identified (by the preambles of the packet) and the corresponding part of the waveform is stored. Then, we recover the waveform to extract the OFDM symbols, i.e., the \gls{csi}. Each beacon contains 52 subcarriers, each carrying one complex symbol. A Python script processes this data by organizing multiple consecutive beacons into a 2D matrix (beacons × subcarriers). Then, the real and imaginary parts of the complex values are separated into two distinct channels, making the data suitable for machine learning classification.

\textbf{mmWave.}
We capture \gls{fpbt} signals with an OpenWRT-modified Mikrotik router and the help of in-house developed Python and bash scripts. This custom firmware modification allows for deep access to the signal processing chain, enabling the extraction of detailed channel state information. More details about this implementation are available in \cite{waveslam}.

At the mmWave band, the MikroTik device sends the output of the \gls{fpbt} to the raw processing block; from this data, the latter extracts I/Q components and received signal power for every antenna of the array, using a multipath signal decomposing algorithm, such as \cite{mdtrack}. The measurements are taken over a full 60-degree beam sweep with given angular resolution so that \gls{pdp} could be calculated for every \gls{awv} configuration. The output of the raw block is the \gls{pdp} for each antenna for each \gls{awv} that forms a particular beam.

We take the raw \gls{csi} data and perform feature extraction. The extracted features include subcarrier I/Q components and \gls{snr}. If the system is in a training mode, we label the data in this step. Labeling is one of the most expensive processes related to \gls{ml}~\cite{liang2022creatingData}, mostly because of the need for human interaction. This fact motivates research on how to automatize the process~\cite{Huy2022BetterDataLabelling} or generate new labelled data~\cite{Leon2018RenderGAN}. 

Hence, we propose using an automatic labeling system, based on YOLOX~\cite{yolox2021}. The mentioned system uses the power of this well-known computer vision ML model to detect the presence of a person. This makes it possible to auto-label all the raw data. Once the training process ends, the camera is removed, preserving people's privacy.

Once the \gls{psta} captures a signal such as a \gls{mmwave} \gls{fpbt} or a 5~GHz beacon the data collection block selects the \gls{csi} and forwards it to the raw processing block, which extracts the relevant features used by \ourmodel{} to classify the scenario.

\subsection{Use cases enabled}

In this work, we demonstrate the capabilities of the proposed prototype through two representative use cases: laboratory presence monitoring and corridor movement tracking. These scenarios have been designed to validate the system’s ability to detect and interpret human activity in different indoor environments. The laboratory presence monitoring use case focuses on identifying whether a person is present within a designated workspace, while the corridor movement tracking use case aims to capture and follow the trajectory of individuals walking along a hallway. The results and analysis of these use cases are presented in detail in the Section \ref{sec:results}.

\bigskip

\textbf{Protoype considerations}

We have tested the performance under interference in the 5~GHz band, which led to a reduced number of usable beacons and a slight decrease in periodicity. In the mmWave band, interference occasionally caused some \gls{fpbt} captures to fail, and these were effectively filtered out during preprocessing. To maintain signal integrity, we incorporated quality checks based on \gls{snr} thresholds and consistency of signal metrics, ensuring that only reliable data is used for sensing.

While our current prototype involves some hardware modifications and a \gls{usrp} to capture 5 GHz signals, we show results that demonstrate the system’s potential to operate robustly even in the presence of interference. With ongoing advancements in commercial Wi-Fi hardware and signal processing techniques, \gls{csi} data can now be extracted from commercial routers, which may support multiple bands and even \gls{mimo} configurations, all accessible from a single device. Thus, we are optimistic that similar performance can be achieved with off-the-shelf devices, making this a practical and scalable solution for real-world passive sensing applications. 

Furthermore, the modularity of our approach allows easy integration with existing infrastructure, reducing deployment costs and complexity. Continuous improvements in \gls{ml} models and adaptive filtering techniques will further enhance system resilience to environmental changes and interference. Overall, these encouraging outcomes highlight a clear pathway toward widespread adoption of passive Wi-Fi sensing in diverse indoor environments.

\section{Experiments and results analysis}
\label{sec:results}

%\begin{figure}
%\centering
%\includegraphics[width=0.95\columnwidth]{IMG/Captura desde 2024-07-15 11-20-56.png}
%\vspace*{0ex}
%\caption{Device capability placeholder}
%\label{fig:Capability}
%\vspace*{-3ex}
%\end{figure}

This section details our experimental testbed, designed to validate the capabilities and applications of multiband passive sensing under the 802.11bf standard. The experiments comprise two test stages: collection time and training; and two use cases: human presence detection and monitoring in their workstation, and human tracking in a corridor.

First, we evaluate our solution capabilities: (i) time spent collecting data, and (ii) training procedure characterization of the model for different values of the training samples and epochs.
Next, we test the system's ability to detect and monitor human presence, placed at different locations. Last, we focus on presence tracking in a corridor. We assess tracking accuracy, providing insights into the system's performance in dynamic environments.

%All data employed to obtain the results is available to facilitate reproducibility and future research\footnote{URL to dataset.}.
%The data is organized in folders for each band, (explain) Also, the code is stored in a public repository\footnote{URL to code.}.

\subsection{Solution Capabilities}
 
\subsubsection{\textbf{Collection time}}
%goal
It is interesting to benchmark if our system is affected by network traffic, as theoretically passive sensing should be independent from communications and its performance does not decrease even in dense scenarios.

%setup, and data collection
To check this, we performed stress tests where we generate different traffic loads. These loads are generated by connecting different \gls{sta}s to the \gls{ap} and introducing traffic using the \emph{iperf} tool. Then, we evaluated the time to collect and process each signal using Matlab. Table~\ref{table:collection_time} summarizes the different loads tested and the results obtained.

%plot explanation
Under low traffic conditions, the collection time remains short, allowing signal capture with minimal delays.
As the traffic load increases to medium levels, there is a slight rise in collection time as it is more difficult for our system to synchronize to the \gls{sta} beacons, however, the accuracy is not affected. Even under high traffic loads, the collection time stays stable, demonstrating the system can work even the \gls{ap} is managing high traffic loads without accuracy reduction. Importantly, across all traffic conditions, the inference latency for each sample remains constant, ensuring that the system processes signals on time.

\begin{table}[ht!]
\centering
\caption{Time it takes for our implementation to passively collect and process a signal, depending on the traffic loads and their effect on 5~GHz, mmWave, and inference.}
\label{table:collection_time}
\begin{tabular}{lcccc}
\toprule
\textbf{Traffic Load/ Time (s)} & \textbf{5 GHz} & \textbf{mmWave} & \textbf{Inference}  \\
\midrule
\textbf{Low 50 Mbps} & 0.52 & 0.9 & 0.01 \\
\textbf{Mid 300 Mbps} & 0.74 & 1.02 & 0.01  \\
\textbf{High 500 Mbps} & 0.83 & 1.04 & 0.01  \\
\bottomrule
\end{tabular}
\vspace{0.3cm}
\end{table}

%discussion and observations
As a result, we can conclude that regardless of our prototype not performing equally under different traffic loads, the system can still work in scenarios with multiple users connected to the \gls{ap}, the time increase on data collection being associated with difficulties of our implementation to synchronize with the different signals and not affecting communications.

Beacons are usually sent out at regular intervals, but if the router is busy with a lot of data traffic, the timing of these beacons can become irregular or they might even be delayed. This makes it harder for the \gls{sdr} to pick up on them, because it might miss some beacons or have to sift through more noise and data signals to find them. In summary, the more traffic the Wi-Fi router is handling, the more crowded the wireless channel becomes, making it more difficult for an \gls{sdr} to detect the Wi-Fi beacon.

\subsubsection{\textbf{Training characterization}}
%goal
It is also interesting to measure the cost of model training. As mentioned, collecting and labeling data for training consumes  resources. The cost of this process grows linearly with the amount of samples needed for training. In addition, the energy used during the training is correlated with the employed computing resources, which is also related with the amount samples at training and the number of epochs.

%setup, and data collection
As a result, we evaluate the training process varying two different parameters, the samples used and the number of epochs spent. We perform this test to classify 4, 8, and 16 labels. For the sample test, the number of epochs was kept constant at 100, while for the epoch test, we fixed the number of training samples at 80\%. These values were selected based on common practices in the state of the art and were found to work well for our particular model configuration.

\begin{figure}[b]%
    \subfloat[]{%
        \includegraphics[width=%
        .5\columnwidth]{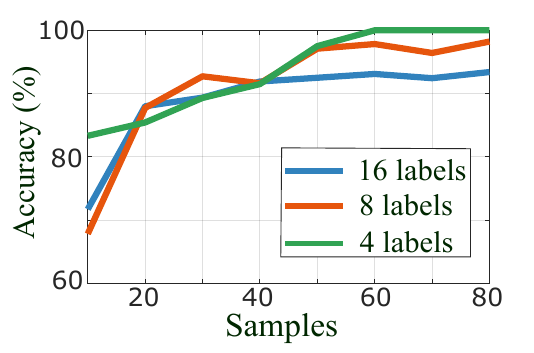}%
        \label{fig:modelSamples}%
    }%
    \subfloat[]{%
        \includegraphics[width=%
        .5\columnwidth]{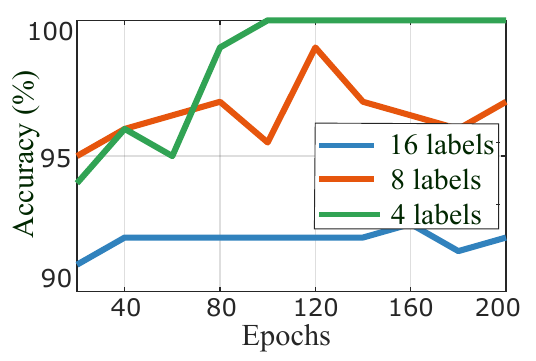}%
        \label{fig:modelEpochs}%
    }%
    \caption{Characterization of the training process of the \gls{cnn} model, varying the number of training samples~(a), and epochs~(b).}
    \label{fig:training}
\end{figure}

%plot explanation + discussion and observations
Fig.~\nref{fig:training}{fig:modelSamples} shows the average accuracy of the model per samples used during training, when the model classify 4 (green), 8 (orange), and 16 (blue) labels. We can see that the accuracy saturates around 60 samples, and it remains stable from there on.

Fig.~\nref{fig:training}{fig:modelEpochs} represents accuracy reached by \ourmodel{} depending on the epoch spent training for the different number of labels mentioned before. As the graph shows, increasing the number of epochs during training improves accuracy up to a certain point, after which additional epochs do not further increase accuracy, in some cases even decreasing it, due to overfitting. This behavior can vary depending on the class or task complexity, but we can agree that around 120 epochs appears to be a good compromise, representing a local maximum in accuracy and likely a near-optimal training point.

\subsection{Indoor people presence monitoring}

%\begin{figure}
%\centering
%\includegraphics[width=0.95\columnwidth]{IMG/Captura desde 2024-07-15 12-46-38.png}
%\vspace*{0ex}
%\caption{sub6 and mmwave accuracies for people detection}
%\label{fig:roomon}
%\vspace*{-3ex}
%\end{figure}

%\begin{figure}
%\centering
%\includegraphics[width=0.95\columnwidth]{IMG/Captura desde 2024-07-15 12-49-17.png}
%\vspace*{0ex}
%\caption{Results pose recognition}
%\label{fig:pose}
%\vspace*{-3ex}
%\end{figure}

\begin{figure}[t]%
    \subfloat[]{%
        \includegraphics[width=%
        \columnwidth]{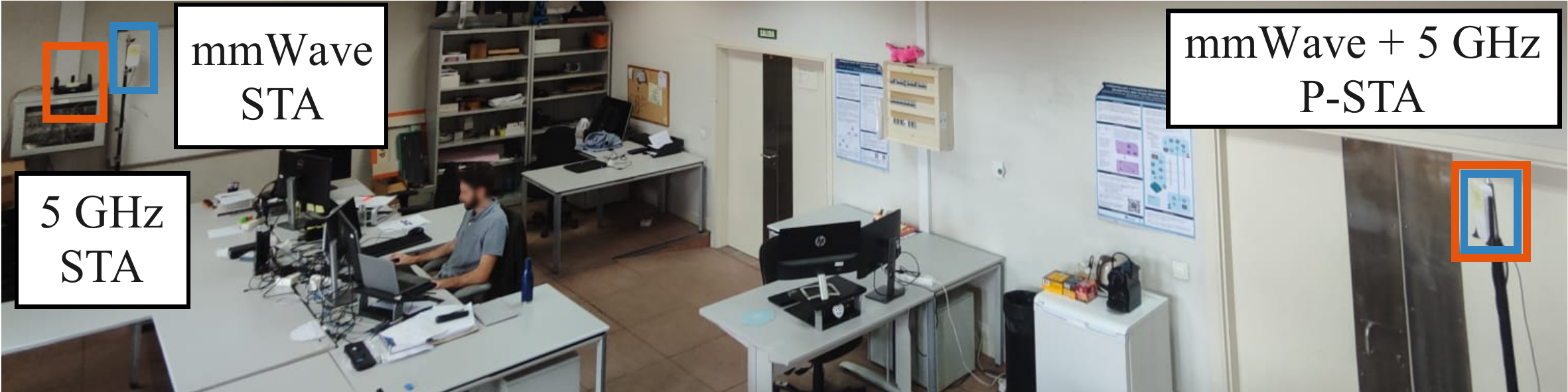}%
        \label{fig:labPicture}%
    }\\%
    \subfloat[]{%
        \includegraphics[width=%
        .55\columnwidth]{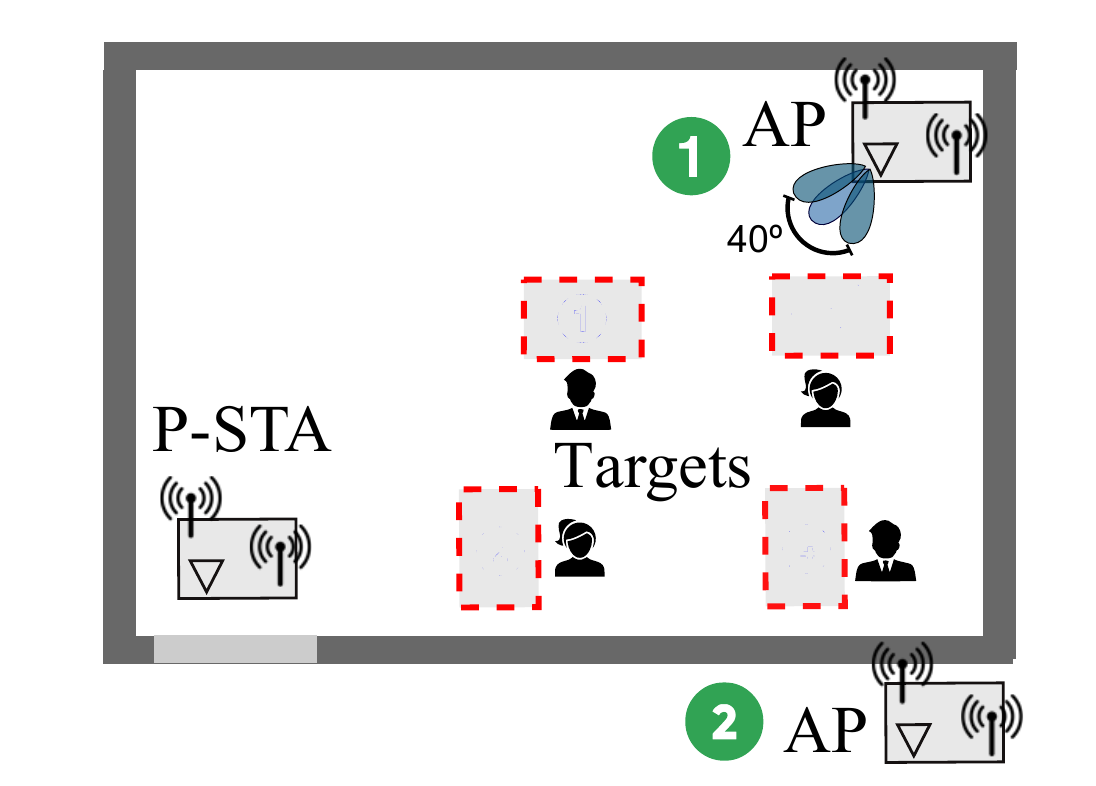}%
        \label{fig:scheme}%
    }%
    \subfloat[]{%
        \includegraphics[width=%
        .45\columnwidth]{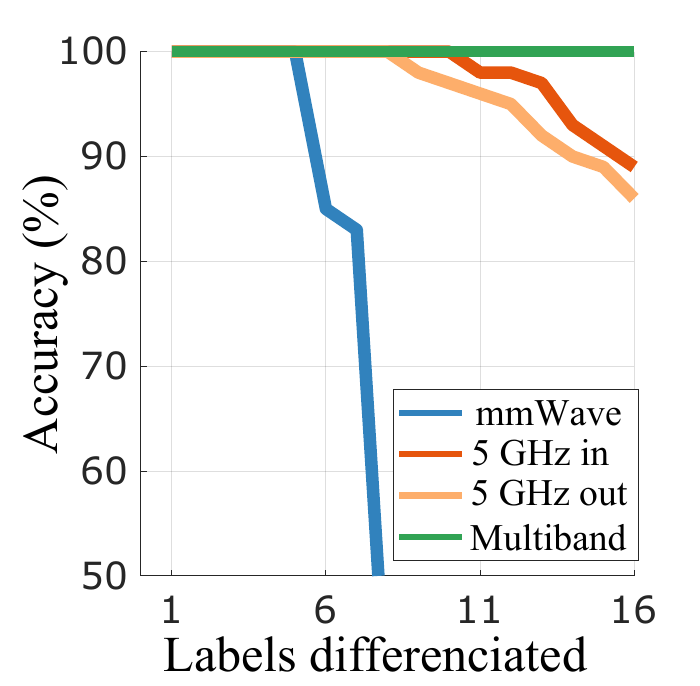}%
        \label{fig:res}%
    }%
    \caption{Indoor people monitoring scenario picture (a), schematic using AP inside the lab-(1) and existing AP outside the lab-(2) (b) and results (c).}
    \label{fig:labResults}
\end{figure}

To test the capabilities of \ourmodel{}, we evaluate it in a real-world scenario. The test involves detecting individuals seated at their designated workstations within a research lab setting. Both 5~GHz and mmWave \gls{ap}s are positioned in one corner of the lab, and data is collected from the opposite corner in a first experiment, and from the corridor outside the room in a second one, depicted in Fig.~\nref{fig:labResults}{fig:scheme}. For this first experiment, participants were allowed to sit at any available desk. In addition, the testing points were evenly distributed across the four classes for both seats and gestures, being each gesture repeated 100 times for each position.

The objective of this experiment is to assess the effectiveness of the multiband passive sensing system in accurately detecting the presence of individuals at specific locations within the research lab, as well as detecting their position.

\subsubsection{\textbf{\gls{ap} inside the room}}

%experimental setup and flow
Firstly, we place both active \gls{ap} and \gls{psta} in the same room, as shown in \nref{fig:labResults}{fig:labPicture}. We evaluate people on four different lab workstations, resulting in 16 different scenarios to differentiate, see \nref{fig:labResults}{fig:scheme}. We perform this experiment for both 5~GHz and mmwave independently. Then we join information from both bands to compare the accuracy achieved when operating together.

%plot
As Fig.~\nref{fig:labResults}{fig:res} shows, using only 5~GHz, the accuracy decreases with the complexity of the scenario, and becomes 89\% when considering all the possible labels. When only using mmWave, cases where the target does not interfere with the \gls{los} path are not properly differentiated, resulting in wrong identifications. Still, mmWave accuracy remains 100\% differentiating if there are \gls{los}, which allows us to detect if a determined path has been obstructed and reduce the search performed by the model using the 5~GHz data. As a result, when joining both sources of information we can achieve 100\% accuracy labeling the 16 scenarios. 

It is worth mentioning that even when the model is trained to detect a particular person, it can successfully detect a different individual without any decrease in performance. In our case, the model was trained with up to four different people and tested on up to 5 different individuals.

\begin{figure}[t]%
    \subfloat[]{%
        \includegraphics[width=%
        .49\columnwidth]{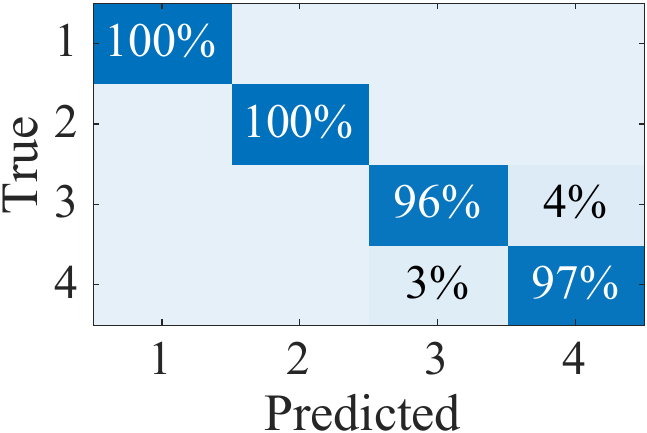}%
        \label{fig:p1}%
    }\hspace{2pt}%
    %\hspace{1.25em}
    \subfloat[]{%
        \includegraphics[width=%
        .49\columnwidth]{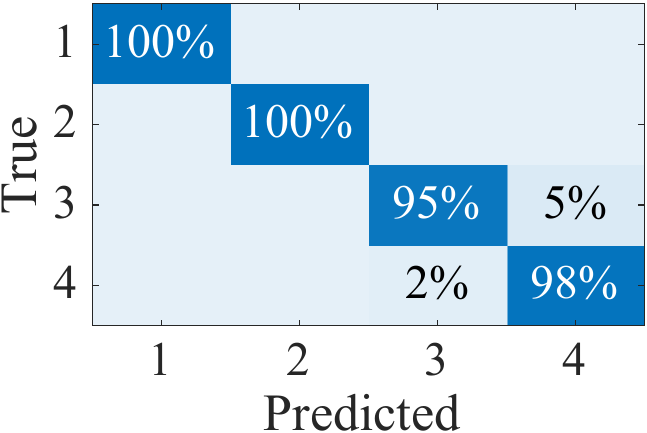}%
        \label{fig:p2}%
    }\vspace{-.5em} \\%
    \subfloat[]{%
        \includegraphics[width=%
        .49\columnwidth]{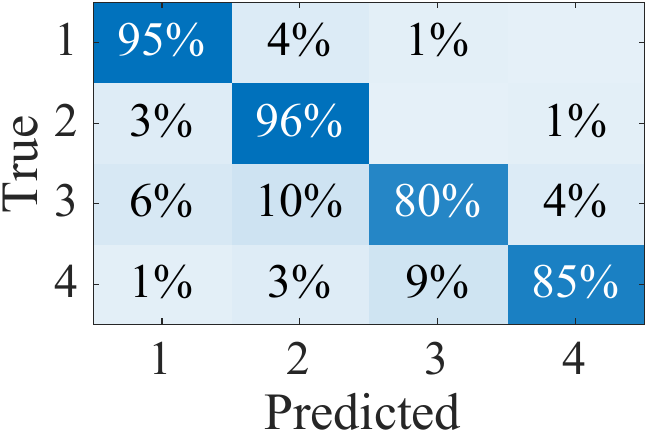}%
        \label{fig:p3}%
    }\hspace{2pt}%
    %\hspace{1.25em}
    \subfloat[]{%
        \includegraphics[width=%
        .49\columnwidth]{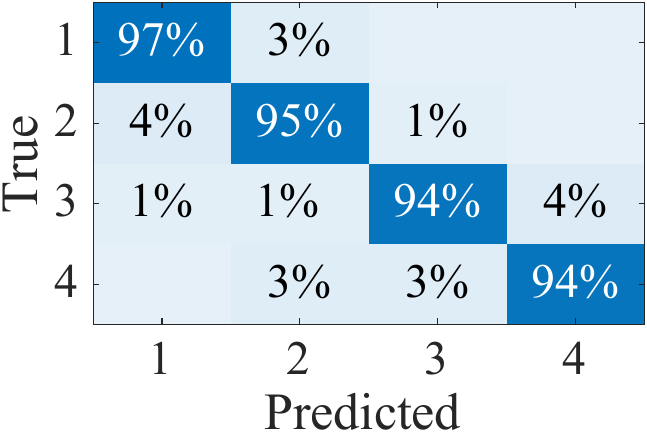}%
        \label{fig:p4}%
    }%
    \caption{Confusion matrices of activity recognition for the different workstations (a-d). Numbers do indicate the status: sit (1), stand (2), lying down (3), and missing (4).}
    \label{fig:pose}
\end{figure}

%plot2
Going further, we evaluate the capabilities of \ourmodel{} to differentiate the pose of a target. We differentiate four poses of a target: sit (1), stand (2) lying down (3), or missing (4). Fig.~\ref{fig:pose} shows the confusion matrices of the different workstations. The accuracy achieved is good for all four workstations, however, those that are further from the \gls{ap}, experience less accuracy detecting scenarios as lying down or missing.

\subsubsection{\textbf{\gls{ap} in the corridor}}
%experimental setup and flow
In this experiment, we use a pre-existing \gls{ap}, already deployed in a corridor, outside the lab. As they only use sub-7~GHz bands, we can just collect that information. In addition, mmWave would be obstructed by the walls of the room, becoming unusable for this particular scenario. Thus, we repeat the same collection process explained previously and compute the accuracy achieved with this approach.

%plot
Fig.~\nref{fig:labResults}{fig:res} shows that, despite the \gls{ap} being outside, we are still able to achieve good results, that is lower than joining both bands' information but could work for particular use cases.

\bigskip

As an additional measurement, we extracted the confidence of the system's decisions, which expresses how reliable is the classification made by \ourmodel{}. 
The average confidence using only 5~GHz is 95.76\% and 92.96\%, when the \glspl{ap} is located inside the room or in the corridor, respectively. However, it grows to 99.92\% when we employ the joint mmWave and 5~GHz solution.

%\vspace{-0.5cm}
\subsection{Corridor movement tracking}
\label{subsec:corridor}

%goal of the experiment
Tracking people in a corridor using only Wi-Fi signals can be highly beneficial in various scenarios, particularly for enhancing security, monitoring, and efficiency in environments such as office buildings, hospitals, or schools. By leveraging existing Wi-Fi infrastructure, we aim to identify the movement of individuals over a corridor and determine what room are they entering.

\begin{figure}[t]
    \centering
    % First figure (half of the column height)
    \subfloat[]{\includegraphics[width=0.47\columnwidth]{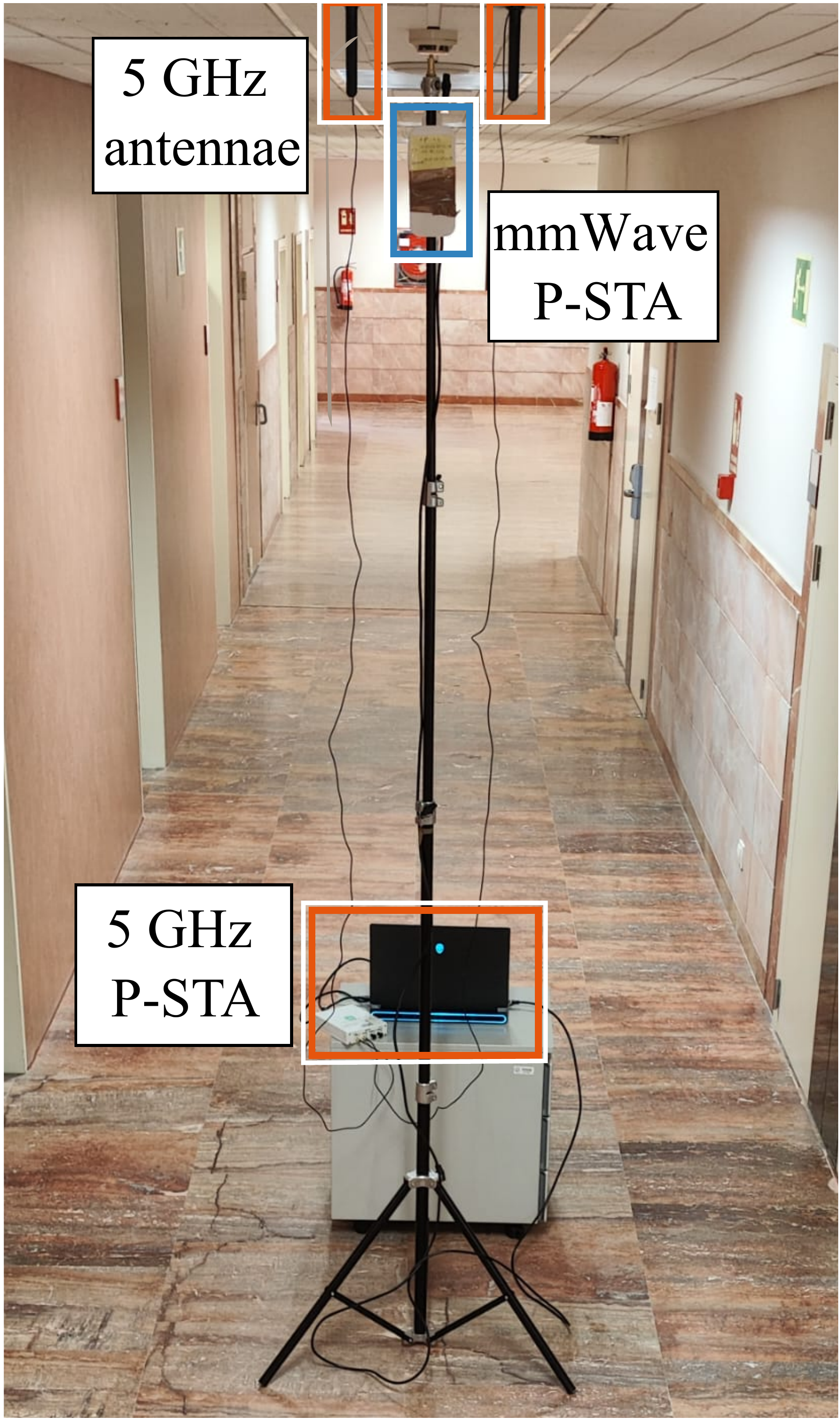}\label{c1}}
    \begin{minipage}[b]{0.52\columnwidth}%
    \centering
    \subfloat[]{\includegraphics[width=\columnwidth]{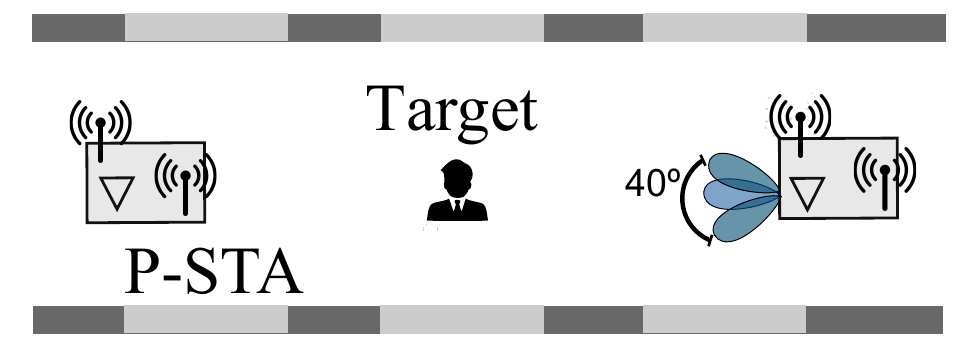}\label{c2}}\\
    \vspace{-.4cm}
    \subfloat[]{\includegraphics[width=.9\columnwidth]{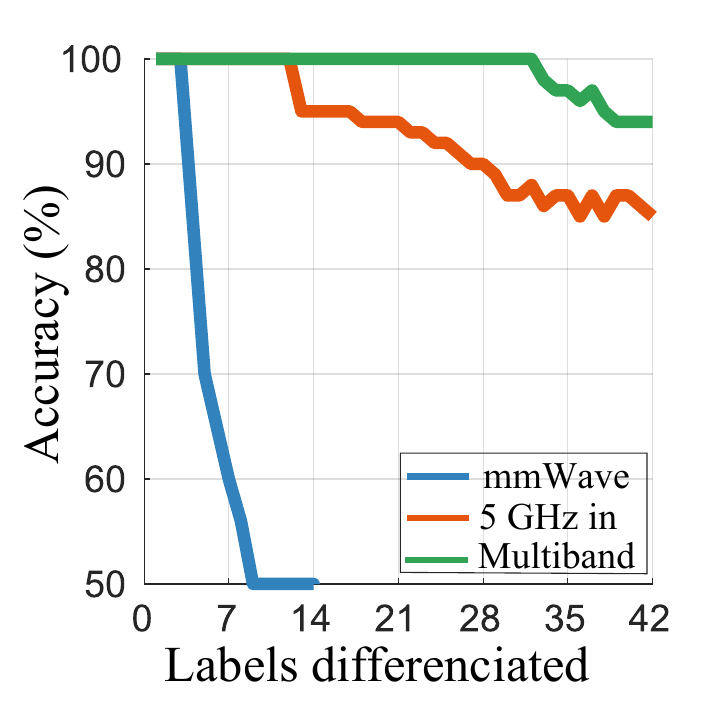}\label{c3}}
    \vspace{.2cm}
    \end{minipage}
    \caption{Corridor presence tracking scenario picture (a), schematic (b) and results (c).}
    \label{fig:expcorridor}
\end{figure}

\begin{figure}[b]
\centering
\includegraphics[width=0.95\columnwidth]{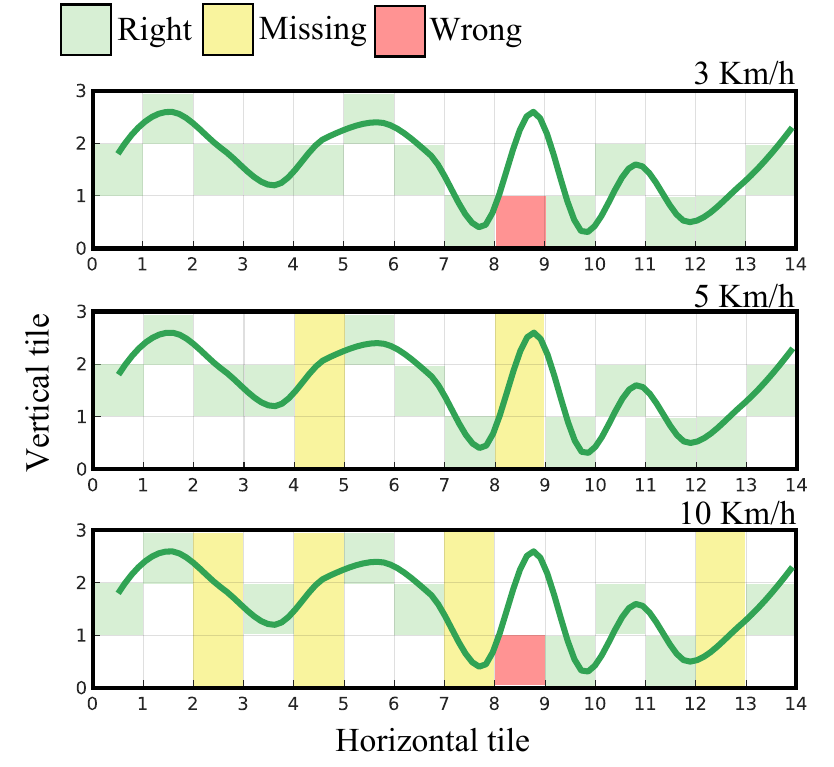}
\vspace*{0ex}
\caption{Results on corridor movement monitoring depending on the speed of the tracked user}
\label{fig:corridor1}
\vspace*{-3ex}
\end{figure}

This enables non-invasive tracking without relying on additional infrastructure like cameras or dedicated sensors, thereby preserving privacy while still collecting meaningful data. For instance, such tracking can support access control to restricted areas, monitor room occupancy to ensure compliance with safety regulations, or log entry times for attendance and scheduling. While our experiments were conducted using a USRP device, which provides high-quality channel measurements, we emphasize that recent efforts have demonstrated the feasibility of extracting \gls{csi} from low-cost commercial Wi-Fi routers. These advances suggest that practical, low-cost deployments are increasingly viable, provided suitable hardware is available \footnote{\url{https://github.com/weisgroup/Openwrt-Atheros-CSI-Tool-for-Netgear-WNDR3700v4}} Wi-Fi-based tracking thus remains a cost-effective solution, especially in scenarios where institutions can leverage existing infrastructure and compatible chipsets to monitor movement and manage space usage efficiently.

%experimental setup and flow
In this experiment, we installed a mmWave device on top of an existing 5 GHz \gls{ap} within the university infrastructure. Then we evaluated the performance of \ourmodel{} in detecting people moving through a corridor (see Fig.~\nref{fig:expcorridor}{c1}). The results show that our system can accurately detect the path taken by individuals by identifying the specific floor tiles they step on, which in turn allows us to infer which rooms they visit.

We evaluated the accuracy of our system in this scenario, where the \gls{psta} collects information on existing signals in a corridor, as Fig.~\nref{fig:expcorridor}{c2} represents. We chose a corridor fragment that has 11.2~m length (14~tiles) and 1.8~m width (3~tiles). As Fig.~\nref{fig:expcorridor}{c3} shows, 5~GHz beacons offer a good accuracy when differentiating up to 14 labels, but is reduced to 85\% when considering all labels. This accuracy is increased back to 94\% when combining 5~GHz band with mmWave information.

In addition, we test the detection of a random path on the corridor depending on different walker speeds: 3, 5 and 10~Km/h. For this purpose, we combine both 5~GHz and mmWave data. mmWave data filters in which vertical tile is the target through beam sweeping, while 5~GHz determines the horizontal position of the user.

%explanation of the plot
Fig.~\ref{fig:corridor1} shows the accuracy of detecting each position depending on the user speed. For 3~Km/h the user is detected correctly in 13 out of 14 positions and the system does not miss any tile. When the user is moving at 5~Km/h, the system misses two tiles, estimating correctly 12 out of 14. However, when we increase the speed of the user, the system misses more tiles. At 10~Km/h, 4 out of 14 tiles are lost and one is wrongly estimated.

\subsection{Spatial and Temporal Generalization}
\label{subsec:generalization}

To evaluate the robustness of our system beyond the original training conditions, we consider two key aspects: spatial generalization, i.e., the ability of the model to operate in different physical layouts or rooms, and temporal generalization, i.e., its performance stability over time.

\subsubsection{Spatial Generalization}
This experiment assesses how changes in the environment impact classification accuracy. Two scenarios are considered:

\textbf{Furniture modifications}: After initial training in a fixed room layout, we rearranged elements such as desks, chairs, and added or removed objects. We then evaluated the model without retraining. We repeated this for minor modifications (ie., moving carton boxes), mid modifications (ie., moving filing cabinets), and major modifications (adding extra furniture).

\textbf{Scenario change}: The model was trained in Room A and tested in Room B, which has similar dimensions but different wall materials and furniture.

\subsubsection{Temporal Generalization}
This experiment analyzes the model's performance over time without retraining. We deployed a trained model and repeated the same test scenarios after 6 months, under minimal or no changes in the room layout. The goal was to determine if environmental drift (e.g., dust, minor object displacement, seasonal humidity changes) affects \gls{csi} features and \ourmodel{} performance.

\bigskip

\begin{figure}[hb]
\centering
\includegraphics[width=0.95\columnwidth]{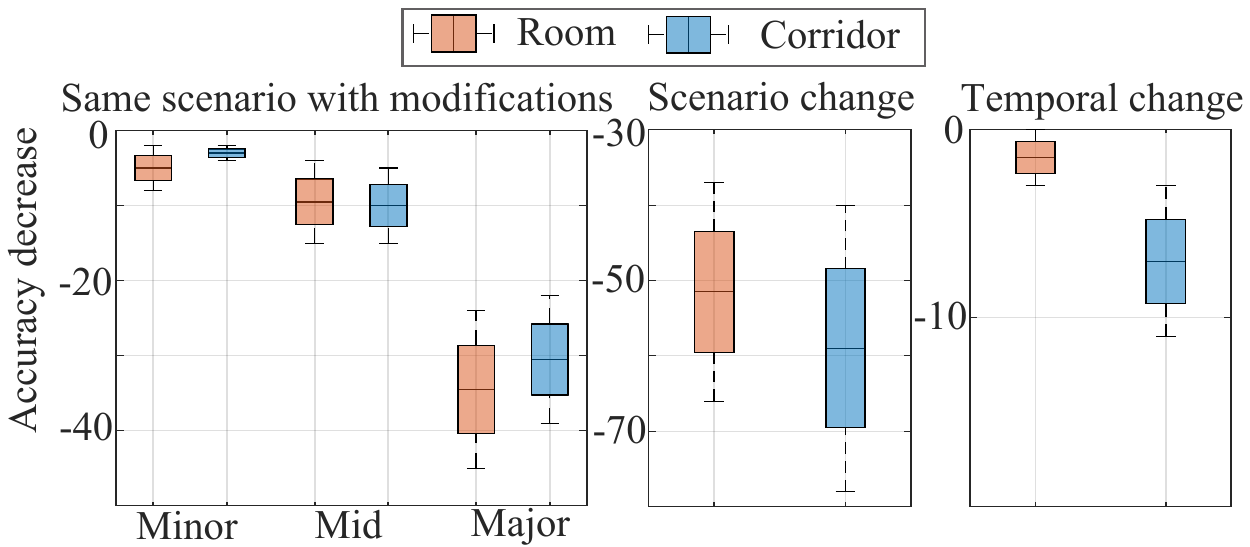}
\vspace*{0ex}
\caption{Accuracy decrease when minor/mid/major modifications are done within the same scenario (left), scenario change (middle), and temporal changes on the same scenario (right)}
\label{fig:changes}
\vspace*{-2ex}
\end{figure}

Results in Fig. \ref{fig:changes}, show that for the same scenario (left), minor or mid modifications do not critically affect the accuracy, but major modifications start to decrease the accuracy up to more than 40\%. For scenario changes (middle), results show that the model does not work and should be retrained with new data. For temporal changes (right), the model seem to work good in a room but slightly decrease the accuracy in corridors (probably due to modifications of rooms next to the corridor, which we cannot control).

As a result, we can define four performance categories:

\begin{itemize}
    \item \textbf{Unchanged Performance}: The model maintains its accuracy with no noticeable degradation (accuracy drop less than 5\%).
    \item \textbf{Minor Degradation}: The model experiences a slight decrease in accuracy between 5\% and 10\%.
    \item \textbf{Moderate Degradation}: The model shows a noticeable drop in accuracy between 10\% and 20\%.
    \item \textbf{Failure}: The model fails to operate as intended, with an accuracy drop exceeding 20\% or complete malfunction (unable to perform the sensing task).
\end{itemize}

\begin{table}[ht!]
\centering
\caption{Sensing accuracy degradation under different spatial and temporal changes}
\label{tab:degradation}
\begin{tabular}{lcc}
\toprule
\textbf{Change Type} & \textbf{Room} & \textbf{Corridor} \\
\midrule
\textit{Same scenario with:}& & \\
Minor modifications    & Minor degradation & Same performance  \\
Mid modifications           &  Minor degradation & Minor degradation  \\
Major modifications      & Critical  & Critical  \\
\midrule
Scenario change       & Failure  & Failure  \\
\midrule 
Temporal change & Unchanged performance & Minor degradation \\
\bottomrule
\end{tabular}
\vspace{0.3cm}
\end{table}
\noindent
\noindent

Table~\ref{tab:degradation} summarizes the impact of spatial and temporal changes on sensing accuracy for the two tested scenarios: a room and a corridor. The main observations are summarized as follows:

\begin{itemize}
    \item \textbf{Minor Modifications}: Small changes resulted in \textit{Minor Degradation} for the room, while the corridor showed \textit{Unchanged Performance}. This suggests that the corridor's simpler geometry makes it less sensitive to slight variations.
    
    \item \textbf{Mid Modifications}: Moderate changes led to \textit{Minor Degradation} in both scenarios. The model retains some generalization capability but is partially affected by altered multi-path conditions.
    
    \item \textbf{Major Modifications}: Significant rearrangements caused \textit{Critical} degradation in both cases, as the severe alteration of the environment's propagation characteristics invalidated the training conditions.
    
    \item \textbf{Scenario Change}: Training and testing in different locations led to \textit{Failure}, highlighting the model’s limited ability to generalize to entirely new environments without retraining.
    
    \item \textbf{Temporal Change}: When evaluated after six months under minimal physical modifications, performance was largely stable, with \textit{Unchanged Performance} in the room and only \textit{Minor Degradation} in the corridor. This indicates good temporal robustness of the model.
\end{itemize}

\bigskip

To summarize, we can extract this set of conclusions on \ourmodel{}:
 
We experimentally confirmed that the performance of the trained model degrades over time when environmental changes occur, such as furniture reconfigurations, but not specifically due to the mere passage of time.

A model trained for a specific space cannot be directly reused in a different room without retraining. This is due to variations in multipath propagation and signal reflections caused by differences in room geometry and materials, which significantly affect CSI-based model performance. Therefore, adaptation to each specific deployment environment is required for reliable operation.

In contrast, the system does generalize well across different individuals. We did not observe significant performance degradation when changing the target person in the same environment, suggesting that the model captures spatial and motion patterns that are robust to inter-person variability.

%PARRAFO NO COMPARISON

\bigskip

%Several significant differences in experimental conditions and methodological decisions make a direct quantitative comparison with the state-of-the-art approaches unrealistic. Different signal types (e.g., ambient Wi-Fi vs. CSI), hardware configurations (e.g., different antenna configurations, use of USRPs or COTS devices), and processing pipelines (e.g., cross-ambiguity functions, spectrograms, machine learning models) are used in each solution. Furthermore, environmental elements like crowd dynamics, room layout, and interference conditions are frequently out of control and differ greatly between testbeds. Standardized benchmarking is impractical due to the biases introduced by these variations. Instead, our work should be seen as a complementary contribution that expands the design space beyond what is currently available in single-band or single-purpose systems by introducing a unified multiband CSI-based framework optimized for low-complexity, real-time execution.

\section{Comparison with State-of-the-art  solutions}

To provide a meaningful evaluation, we implemented two representative single-band solutions from \gls{soa}: one operating in Wi-Fi 5~GHz \cite{comparisonnew1} and the other in mmWave \cite{comparisonnew2}. These are the two most representative solutions from the recent \gls{soa} that closely align with our setup and objectives. The evaluation is structured as follows. First, we describe the two \gls{soa} methods and then we perform a multiband combination to compare with \ourmodel{}. Last, we assess their performance in two different environments: \textit(i) people monitoring inside a lab, and \textit(ii) person tracking in a corridor.

\subsection{State of the art solutions}

\textbf{Sensor-Aided Learning for Wi-Fi Positioning With Beacon CSI \cite{comparisonnew1}}: The method leverages \gls{csi} from beacon frames, in addition to traditional received \gls{rssi}, to capture the propagation conditions of each access point. This allows the model to adapt to \gls{los} and \gls{nlos} environments, thereby improving positioning accuracy while requiring minimal manual calibration. The data consist of \gls{csi} and \gls{rssi} values extracted from beacon frames received at multiple Wi-Fi access points. In our 5 GHz approach, however, \gls{rssi} is not included, and only \gls{csi} information is used for the analysis.

\textbf{Overhead-Free People Counting in mmWave Networks Using IEEE 802.11bf Passive Sensing \cite{comparisonnew2}}: The paper proposes a method that leverages IEEE 802.11bf passive sensing at mmWave frequencies, where \gls{dmg} beacon frames are used as the basis for environmental sensing. Since these beacons are already part of the standard communication process, the approach claims to introduce no additional overhead and allows for efficient people counting while maintaining low energy consumption. In contrast, our mmWave approach also employs a standard signal but relies on Beam Training (BT) frames rather than \gls{dmg} beacons.

\textbf{Multiband combined solution}: It is important to note that the \gls{soa} solutions described above were each designed for single-band operation, either in Wi-Fi 5 GHz or in mmWave. None of them incorporates a multiband design, which is precisely one of the main contributions of our work.

However, to enable a fairer comparison, we implemented a multiband variant by combining the outputs of both \gls{soa} methods. Specifically, we followed a late-fusion approach adapted following the techniques proposed in \ourmodel{}.

\begin{figure}%
\centering
    \subfloat[]{%
        \includegraphics[width=%
        .5\columnwidth]{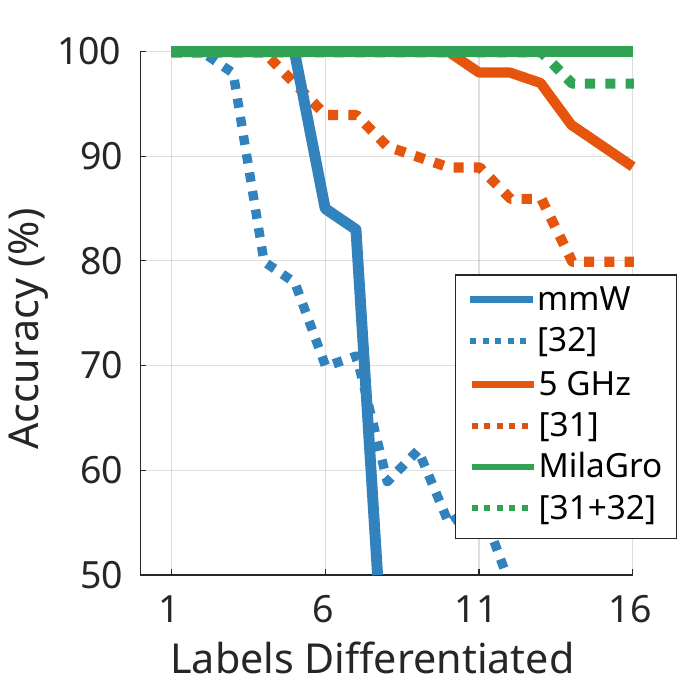}%
        \label{fig:c1}%
    }%
    \subfloat[]{%
        \includegraphics[width=%
            .5\columnwidth]{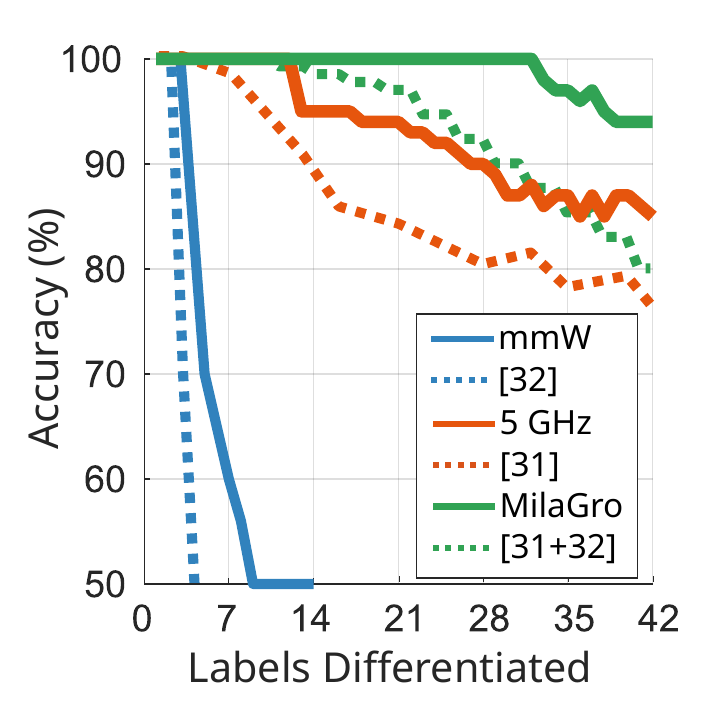}%
        \label{fig:c2}%
    }%
    \caption{Comparison of \ourmodel{} with the \gls{soa} in a indoor lab~(a), and in the corridor~(b).}
    \label{fig:compare}
\end{figure}

\subsection{Experimental evaluation}

We evaluate the performance of our proposed solutions across two different environments: the indoor lab scenario and the corridor scenario (see Fig.~\ref{fig:labResults},Fig.~\ref{fig:expcorridor}). In both cases, we assess the ability of each method to detect or track people under varying numbers of labeled classes. The results are presented for 5~GHz, mmWave, and their multiband combination.

\subsubsection{5 GHz}

As shown in Fig.~\ref{fig:c1}, the performance of the 5~GHz solution is consistent across both scenarios. Our method achieves results comparable to the representative \gls{soa} approach, which aligns with the extensive prior use of 5~GHz Wi-Fi beacon sensing for presence detection and tracking. Accuracy gradually decreases as the number of classes increases, with the \gls{soa} solution dropping in high-class cases. This trend demonstrates that while 5~GHz sensing is reliable, it has inherent limitations when finer classification granularity is required.

\subsubsection{mmWave}

The mmWave results, illustrated in Fig.~\ref{fig:c2}, show more scenario-dependent behavior. In the lab environment, the \gls{soa} method underperforms for a low number of classes but degrades more slowly as the number of classes increases, thanks to its more complex, standalone mmWave design. Our simpler mmWave model performs better for fewer classes but shows a steeper decline with increasing complexity. In the corridor scenario, these limitations become more pronounced: the \gls{soa} solution struggles beyond three classes, causing a drastic drop in accuracy. Our mmWave model exhibits similar trends, performing adequately with low number of classes but deteriorating rapidly once these are increased.

\subsubsection{Multiband}

Combining both frequency bands demonstrates the clear benefits of multiband sensing (Fig.~\ref{fig:compare}). Extending the \gls{soa} methods to a multiband setup improves accuracy relative to single-band operation; however, the fused \gls{soa} solution inherits the high mmWave errors even with low classes, which limits overall performance.

By contrast, our proposed multiband approach, designed from the ground up to integrate sub-7~GHz and mmWave information, consistently outperforms both the single-band and combined \gls{soa} methods across all scenarios and class configurations. This is possible thanks to our model layer structure, which is designed to handle and integrate signals from different bands effectively. As a key contribution of this paper, we show that using multiple bands for passive sensing complements single-band approaches, increasing the sensing accuracy. Effective multiband solutions require dedicated model planning to fully exploit the available frequency diversity, rather than retrofitting existing single-band designs.

\subsection{Conclusions of \gls{soa} comparison}

The results clearly demonstrate the advantages of exploiting multiple frequency bands. While single-band methods provide useful baselines, they quickly face limitations when the classification task becomes more complex. Extending \gls{soa} solutions to a multiband setup confirms that combining different signals is always beneficial, as it improves accuracy compared to operating in isolation. However, these retrofitted approaches remain constrained, since their models were never designed to jointly process heterogeneous inputs.

Our proposed multiband approach, \ourmodel{}, in contrast, was conceived from the outset to integrate 5~GHz and mmWave information, leading to consistently higher accuracy across both monitoring and tracking scenarios. Multiband sensing not only mitigates the weaknesses of each band but also enables robust and scalable performance.

\section{Security considerations}
\label{sec:sec}

The proposed system and prototype have demonstrated a \gls{psta} can sense people with fairly good accuracy in multiple scenarios, without interfering in the communication process and even without registration. This may raise alarms, as it may suppose a privacy leakage. In the following paragraphs, we discuss potential threats that we identified for our system, analyzing possible solutions.

\begin{figure}[b]
    \subfloat[]{%
        \includegraphics[width=%trim={0 0 2.2cm 1.1cm}
        .47\columnwidth]{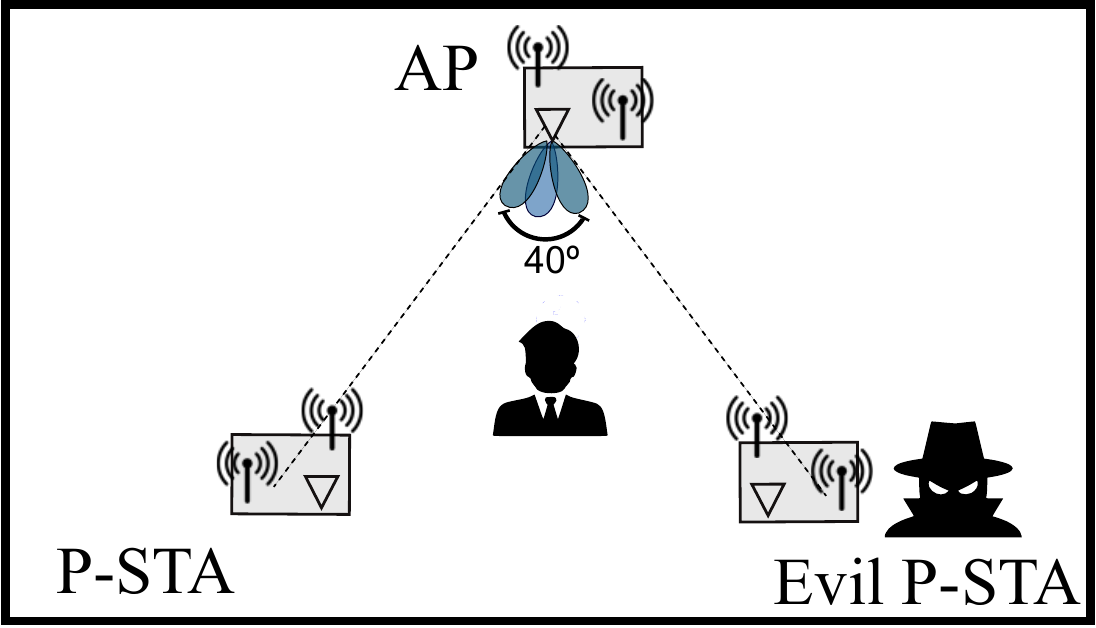}%
        \label{fig:malpsta}%
    }%
    \hspace{.06\columnwidth}%
    \subfloat[]{%
        \includegraphics[width=%
        .47\columnwidth]{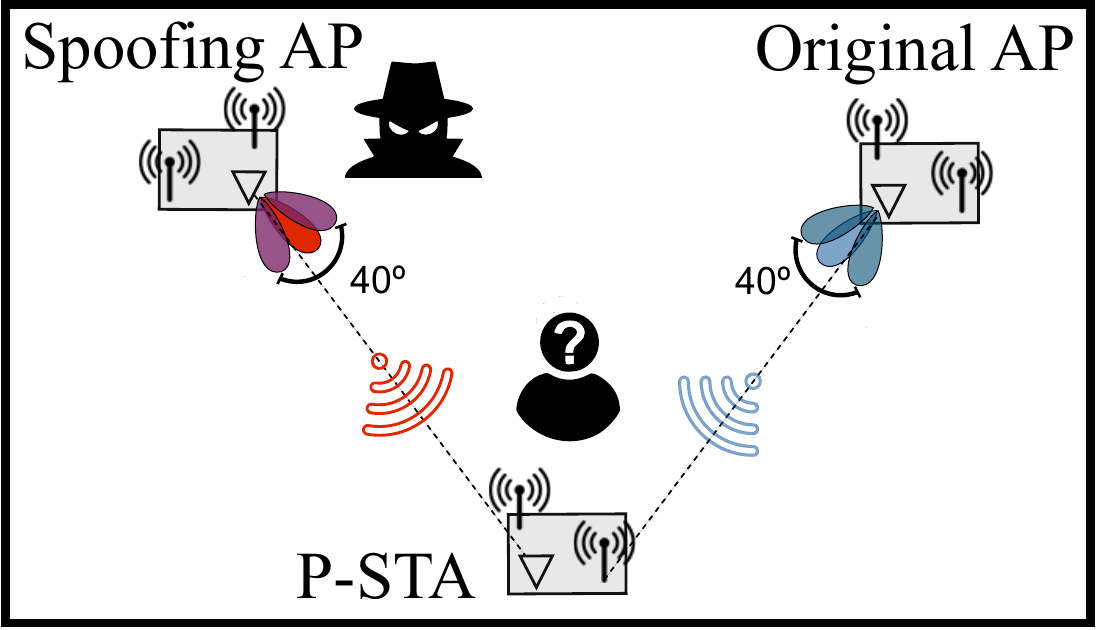}%
        \label{fig:malap}%
    }%
    \caption{Schematic of scenarios where: (a) evil \gls{psta} senses the environment without authorization, or (b) spoofing \gls{ap} interferes in passive sensing by generating undesired interference.}
    \label{fig:securitySchemes}
\end{figure}

Due to the fact that the \gls{psta} is not registered, it is unavoidable that an evil \gls{psta} can sense the environment, as Fig.~\nref{fig:securitySchemes}{fig:malpsta} shows. However, it is mandatory to collect and label data for training the model before starting sensing. This process adds difficulties to malicious \glspl{psta}, which need access to the labeled data. For example, a housebreaker cannot track human presence inside a particular house, as long as it does not have access to that information from other ways for training the model.
Additionally, replicating the target scenario in another place (for labeling there the data) is also not possible, because adjacent rooms, and even cables and pipes under the walls, influence the results. To prove that, we have conducted some experiments. We have employed the model from Section~\ref{subsec:corridor} in a similar corridor, with the same dimensions and wall materials. Even under similar conditions, the model was unable to locate a person, showing an accuracy that fluctuates from 0\% to 10\%. Despite this, when testing the same corridor using a previously trained model—approximately 7 months old—we found no evidence of performance degradation over time.

%However, we have identified potential threats for our system that are discussed below. 

Although a malicious \gls{psta} needs labeled data to locate people, it still can perceive variations in the \gls{csi} over time, which indicates movement. Under the right circumstances, that fact can pose a dangerous threat, for example, when looking for inhabited houses. To prevent it, we can artificially modify the \gls{csi} to simulate movement, for example, altering the router beacon interval or the antenna polarization configuration.

In case the malicious \gls{psta} has access to labeled data, it can take advantage of this to unequivocally identify people without their consent, violating its privacy.

We have performed experiments to probe if our system can distinguish between people, to identify and track them. Despite some articles tackling this problem using \gls{mmwave} antennas, they use different modulations and techniques that provide higher resolution. Our system is not able to recognize people, giving close to random answers, as listening to a beamforming procedure cannot offer that resolution. Unless their physical differences are very significant, for example, an adult versus a kid, the system can sense if there is a person, but cannot identify who is it. This prevents malicious \glspl{psta} from tracking or detecting some particular user, which would pose a privacy threat.

\begin{figure}
    \subfloat[]{%
        \includegraphics[width=%trim={0 0 2.2cm 1.1cm}
        .5\columnwidth]{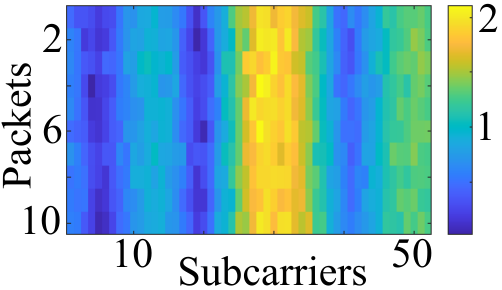}%
        \label{fig:1apstatic}%
    }%
    \subfloat[]{%
        \includegraphics[width=%
        .5\columnwidth]{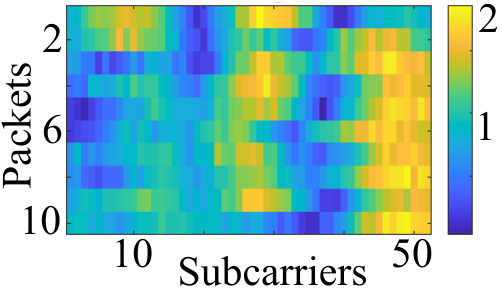}%
        \label{fig:1apmove}%
    }\\%
    \subfloat[]{%
        \includegraphics[width=%
        .5\columnwidth]{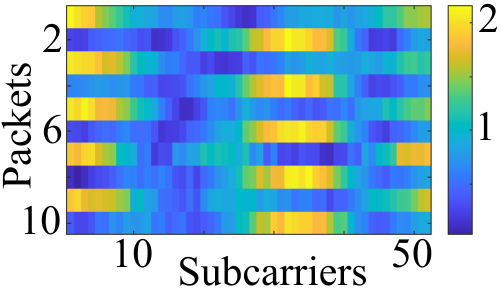}%
        \label{fig:2apstatic}%
    }%
    \subfloat[]{%
        \includegraphics[width=%
        .5\columnwidth]{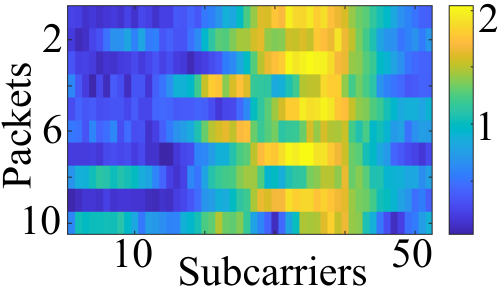}%
        \label{fig:2apmove}%
    }%
    \caption{\gls{csi} from single \gls{ap} (a, b) and adding spoofing \gls{ap} (c, d) in a static and a moving environment, respectively.}
    \label{fig:impersonatingAP}
\end{figure}
Another kind of attack that can affect our system is the {denial of service}, or in other words, blocking the system to force deriving the results by a trusted \gls{sta}. The first way to deny the service is by introducing noise in the same channel, but this will also result in a denial of the communications. By impersonating the \gls{ap} that sends the beacons (spoofing), one can also produce wrong measurements, see Fig.~\nref{fig:securitySchemes}{fig:malap}; notice that the \gls{psta} only listens to trusted \glspl{ap}.

We further investigate this last point through experiments, deploying two \gls{ap}s with the same SSID, band, and channel. Fig.~\ref{fig:impersonatingAP} shows different signals collected by the \gls{psta}. Top figures belong to a scenario with a single \gls{ap} sending beacons, without (Fig.~\nref{fig:impersonatingAP}{fig:1apstatic}) and with (Fig.~\nref{fig:impersonatingAP}{fig:1apmove}) mobility. Adding an extra \gls{ap} implies that the \gls{psta} receives beacons from both of them, corrupting the measurement (Figures~\nref{fig:impersonatingAP}{fig:2apstatic} and~\nref{fig:impersonatingAP}{fig:2apmove}, respectively). Notice that there is a pattern in Fig.~\nref{fig:impersonatingAP}{fig:2apstatic} that can be used to recognize the correct beacons. However, the pattern becomes extremely complex using different techniques, for example, adding more \glspl{ap}, dynamically changing the beacon interval. These insights can be used for malicious \glspl{ap}, to deny the service; or to protect a scenario from malicious \gls{psta}, to avoid being sensed.

Despite the ability of a spoofing \gls{ap} to fake a deliberated labeled result to trick the \gls{psta} and force specific wrong environmental estimations, it is barely possible to generate the same signal as a benignant \gls{ap} from another location, due to the complexity of estimating the signal propagation. In addition, a well-deployed Wi-Fi infrastructure can easily detect a rogue \gls{ap} Hence, we consider that this kind of attack is not a problem for our system.

All these security considerations are related to both sub-7~GHz and \gls{mmwave} bands. Nevertheless, the latter is less susceptible due to its reduced range and increased complexity~-- the \gls{ap} and \gls{psta} should have \gls{los} to work properly. Additionally, attackers will experience extra difficulties in impersonating an \gls{ap} because \gls{mmwave} antennas are highly directional, making it easier to detect the source of the signal.

\section{Related work}
\label{sec:soa}

This section explores the state of the art on sensing with Wi-Fi signals. We begin with sub-7GHz Wi-Fi sensing approaches, used for basic activity detection and people counting. We then review mmWave-based methods that offer higher resolution and enable more fine-grained tracking, such as gesture recognition, though at the cost of shorter range and poor \gls{nlos} performance. Next, we examine passive sensing techniques that reuse existing transmissions to detect occupancy and movement without active user involvement. Last, we compare our testbed with other state-of-the-art testbeds, which fuse sub-7GHz and mmWave data to enhance robustness and accuracy in real-time occupancy and flow monitoring tasks.

%ra based solutions

%sub6
<%\begin{table*}
%\centering
%\caption{Comparison of State-of-the-Art Testbeds of Wi-Fi Sensing for people tracking/monitoring}
%\label{tab:methods_comparison}
%\begin{tabular}{p{3.2cm} p{3.5cm} p{5.5cm} p{4cm}}
%\toprule
%\textbf{System} & \textbf{Sensing Basis} & \textbf{Signal Processing Technique} & \textbf{Counting Focus} \\
%\midrule
%Passive Wi-Fi Radar \cite{comparison1}& 
%Ambient Wi-Fi signals (not CSI) & 
%Cross Ambiguity Function (CAF), Range-Doppler maps, CLEAN algorithm (CNNs) & 
%Occupancy detection and coarse people counting \\
%\midrule
%Wi-PC \cite{comparison3}& 
%Channel State Information (CSI) & 
%Feature extraction, Support Vector Machine (SVM) classifier & 
%People counting (up to 5) in enclosed spaces \\
%\midrule
%WiFlowCount \cite{comparison2}& 
%CSI with Doppler exploitation & 
%Doppler spectrogram construction, optimal segmentation, CNNs  & 
%People flow and subflow counting in passages (not specific path) \\
%\midrule
%\textbf{Our Solution} & 
%Multiband Wi-Fi CSI + PDP of Beam Training & 
%Multiband feature fusion based on machine learning  & 
%Both occupancy detection and people flow counting \\
%\bottomrule
%\end{tabular}
%\vspace{0.3cm}
%\end{table*}

\subsection{Sub-7 GHz Wi-Fi Sensing}
%introduction to sub6
Commodity Wi-Fi devices employing technologies such as IEEE 802.11ac and IEEE 802.11ax, operating at 2.4~GHz and 5~GHz, have demonstrated effectiveness in human sensing tasks, including activity recognition, vital sign monitoring, and person identification.

%low gran
Certain methods leverage low-granularity fingerprint-based approaches, as discussed in \cite{finger,finger1,finger2}, enhancing accuracy through \gls{ml} techniques. While these solutions are cost-effective, they are not as robust as radar-based solutions, as they suffer from errors in positioning of the order of few meters.

%recent signal 
Recent and more sophisticated methods achieve higher accuracy by integrating Multi User \gls{mimo}, \gls{ofdm} transmissions, and analyzing \gls{csi} amplitude across subcarriers \cite{surveyWiFi,WiFivision,efficientfi}. These solutions achieve accuracies from 90\% to 95\%, detecting targets located at few meters. However, these technologies present two primary limitations: they are predominantly effective in single-person scenarios due to bandwidth constraints, leading to coarse localization and tracking, and they exhibit high sensitivity to environmental changes. This sensitivity hinders generalization to new, unseen scenarios, significantly degrading performance.

%limitations and solutions so far
To address these limitations, approaches such as those in \cite{multisense} utilize multiple devices with multiple antennae for monitoring the vital signs of multiple individuals, achieving an accuracy above 90\% for multiple scenarios. Nonetheless, these solutions are specialized and necessitate custom Wi-Fi deployments, diverging from commodity device use. Advanced deep learning and optimization techniques \cite{ml1,ml2} have the potential to reduce environmental dependency by incorporating computational classification layers that enhance accuracy in complex scenarios. Despite these advancements, the field remains in an exploratory phase, with existing solutions largely tailored to specific scenarios.

%mmwave

\subsection{mmWave Wi-Fi Sensing}
%status mmwave
Recently, mmWave radios have been introduced for sensing tasks such as client device localization, people tracking, fine-grained human gesture recognition, and vital sign monitoring. The high resolution and bandwidth of mmWave provide higher precision compared to sub-7~GHz radios \cite{mmwavesur}.

%recent solutions and problems
Solutions like mmSense \cite{mmwavemobi} and RAPID \cite{rapid} can detect multiple people using mmWave IEEE-based radios, achieving accuracies comparable to radars. They leverage beam training on high bandwidth and frequency, extracting Doppler and channel response. However, these implementations are not yet commercially available and require very specific hardware designs to extract such information from the channel.

%general mmwave drawbacks
Additionally, mmWave has a short range and performs poorly in \gls{nlos} situations, making multipath estimation challenging. Recent standardization efforts with 802.11az \cite{az} employ \gls{los} assessment techniques to accurately determine if an STA has \gls{los}, utilizing polarization swapping over different training fields. Some works, such as \cite{alejandro}, attempt to mitigate this issue by integrating sub-7~GHz information into the mmWave data.

\subsection{Passive Sensing}
%passive sensing
It has become an attractive alternative for many applications, allowing the use of signals without requiring active involvement from people or devices being monitored. In the case of Wi-Fi-based passive sensing, several approaches have been developed to use Wi-Fi signals for topics like contact tracing, human detection, tracking, and monitoring crowd movement.

A major use case is human detection, where passive radar systems analyze Wi-Fi signals to sense the presence and movement of people. These systems can work even when Wi-Fi access points are not constantly transmitting data, as shown by \cite{p2}, that reuses Wi-Fi when the network is idle. However, these systems do not offer good accuracy in complex scenarios due to the simplicity of the signals collected.

Passive sensing is also used for monitoring pedestrian flow in public spaces. Wi-Fi probe requests sent by mobile devices are captured and used to estimate the number of people and their movement speeds, helping to understand crowd behavior~\cite{p3}. Similarly, public transit ridership can be monitored using passive Wi-Fi sensing. A system combining neural networks was developed to track the number of passengers on public buses in real-time~\cite{p4}. Despite these systems passively connect Wi-Fi requests, they need active Wi-Fi users to work. In addition, they only offer high-level information about people tracking.

Passive Wi-Fi signaling can also be used as a radar. One system detects people moving behind walls using a passive radar receiver to track their movements based on Doppler shifts~\cite{p5}. Another approach improves on this by detecting not only large movements like walking but also small motions like typing or breathing, showing even greater sensitivity~\cite{p6}. Lastly, real-time tracking of a single target has been achieved using commercial Wi-Fi devices, with algorithms that can accurately estimate movement and position based on signal shifts and angles~\cite{p7}. These approaches offer high accuracy but the bands used can easy be affected by interference, reducing the quality of the signal collected.

To the best of our knowledge, no prior studies have simultaneously exploited both the sub-7~GHz and \gls{mmwave} bands for passive sensing, nor have they demonstrated how these two bands can be combined. Our approach integrates both bands, enabling a robust sensing solution that leverages the superior penetration properties of 5~GHz alongside the high-resolution capabilities of \gls{mmwave}.

Unlike active sensing approaches, our solution requires no additional scheduling or coordination between sensing and communication tasks. Both functionalities coexist seamlessly, without degrading the performance of either, as we demonstrate. This dual-band passive sensing strategy thus represents a significant advancement over existing methods, establishing a new direction in the field.

\section{Conclusions}
\label{sec:conclusions}

In this paper, we present a multiband passive Wi-Fi sensor based on the latest IEEE 802.11bf standard. Our system leverages sub-7~GHz and mmWave beacons to provide accurate environmental sensing. This enables the detection and tracking of individuals’ presence and activities in indoor scenarios, such as our research lab. Moreover, we demonstrate the ability to trace individuals’ movement through corridors and distinguish the specific rooms they enter.

Our solution integrates information from both the 5 GHz and mmWave bands through \ourmodel{}, enabling real-time inference. We show that it outperforms single-band passive sensing solutions from the \gls{soa}, highlighting the need for new multiband models to process data in such approaches.

The results show that by combining multiband data, our system can distinguish between complex indoor scenarios and predict movement trajectories with accuracies ranging from 90\% to 100\%. Fusing sub-7 GHz and mmWave bands enables robust performance in challenging scenarios, where a single-band passive sensing approach would lack the resolution needed.

In addition, we address the security challenges by simulating potential scenarios that could expose vulnerabilities in our system. These include evil \glspl{psta} or spoofing \gls{ap}s that can sense the environment without authorization or alter the performance of our system. We analyze these issues, identifying the risks and offering possible alternatives to mitigate them, ensuring the system remains secure while maintaining its performance and reliability.

Future research directions could explore using other types of signals for sensing purposes, integrating multistatic data to enable real-time reconstruction of larger environments, or using other algorithms to improve inference performance in even more complex scenarios. Additionally, expanding the system to outdoor environments or considering multi-floor buildings would offer new challenges and opportunities for enhancing Wi-Fi-based passive sensing systems. Finally, further exploration of privacy-preserving methods is crucial as sensing technologies become increasingly sophisticated and widespread in everyday applications.

\bibliographystyle{IEEEtran}
\bibliography{IEEEabrv,IEEEsettings,ref}{}

\begin{thebibliography}{10}
\providecommand{\url}[1]{#1}
\csname url@samestyle\endcsname
\providecommand{\newblock}{\relax}
\providecommand{\bibinfo}[2]{#2}
\providecommand{\BIBentrySTDinterwordspacing}{\spaceskip=0pt\relax}
\providecommand{\BIBentryALTinterwordstretchfactor}{4}
\providecommand{\BIBentryALTinterwordspacing}{\spaceskip=\fontdimen2\font plus
\BIBentryALTinterwordstretchfactor\fontdimen3\font minus \fontdimen4\font\relax}
\providecommand{\BIBforeignlanguage}[2]{{%
\expandafter\ifx\csname l@#1\endcsname\relax
\typeout{** WARNING: IEEEtran.bst: No hyphenation pattern has been}%
\typeout{** loaded for the language `#1'. Using the pattern for}%
\typeout{** the default language instead.}%
\else
\language=\csname l@#1\endcsname
\fi
#2}}
\providecommand{\BIBdecl}{\relax}
\BIBdecl

\bibitem{WiFi_2024status}
\BIBentryALTinterwordspacing
{Wi-Fi Alliance}, ``The state of connectivity: {Wi-Fi}® momentum in 2024,'' {30/05/2024}. [Online]. Available: \url{https://www.wi-fi.org/beacon/the-beacon/the-state-of-connectivity-wi-fi-momentum-in-2024}
\BIBentrySTDinterwordspacing

\bibitem{passub61}
F.~Filippini \emph{et~al.}, ``{OFDM} based {WiFi} passive sensing: a reference-free non-coherent approach,'' in \emph{IEEE Radar Conference (RadarConf23)}, 2023, pp. 1--6.

\bibitem{passub62}
Z.~He \emph{et~al.}, ``A robust {CSI}-based {Wi-Fi} passive sensing method using attention mechanism deep learning,'' \emph{IEEE Internet of Things Journal}, vol.~10, no.~19, pp. 17\,490--17\,499, 2023.

\bibitem{pasmmwave1}
C.~Yu, Y.~Sun, Y.~Luo, and R.~Wang, ``{mmAlert}: {mmWave} link blockage prediction via passive sensing,'' \emph{IEEE Wireless Communications Letters}, vol.~12, no.~12, pp. 2008--2012, 2023.

\bibitem{pasmmwave2}
J.~Li \emph{et~al.}, ``Passive motion detection via {mmWave} communication system,'' in \emph{IEEE 95th Vehicular Technology Conference: (VTC2022-Spring)}, 2022, pp. 1--6.

\bibitem{3gpp}
{3GPP}, ``Study on 6g use cases and service requirements,'' 3rd Generation Partnership Project (3GPP), TSG SA1, Technical Report (Draft) TR 22.870, 2024, stage-1 expected completion by March 2026.

\bibitem{etsi}
{ETSI ISG ISAC}, ``Integrated sensing and communications (isac); channel modelling, measurements and evaluation methodology,'' European Telecommunications Standards Institute (ETSI), Sophia-Antipolis, France, Group Report GR ISC 002 V1.1.1, Aug. 2025, 62 pages.

\bibitem{bf}
{IEEE Task Group BF}, ``{IEEE} draft standard for information technology -- telecommunications and information exchange between systems local and metropolitan area networks -- specific requirements - part 11: Wireless {LAN} {M}edium {A}ccess {C}ontrol ({MAC}) and {P}hysical layer ({PHY}) specifications amendment 4: Enhancements for wireless {LAN} sensing,'' \emph{IEEE P802.11bf/D3.0, December 2023}, pp. 1--227, 2024.

\bibitem{Cao2021}
Y.~Cao \emph{et~al.}, ``A lightweight deep learning algorithm for {WiFi}-based identity recognition,'' \emph{IEEE Internet of Things Journal}, vol.~8, no.~24, pp. 17\,449--17\,459, 2021.

\bibitem{WiFi-Sleep}
B.~Yu \emph{et~al.}, ``{WiFi-Sleep}: Sleep stage monitoring using commodity {Wi-Fi} devices,'' \emph{IEEE Internet of Things Journal}, vol.~8, no.~18, pp. 13\,900--13\,913, 2021.

\bibitem{PLmodel_sub6}
T.~Adame, M.~Carrascosa, and B.~Bellalta, ``The {TMB} path loss model for 5 {GHz} indoor {WiFi} scenarios: On the empirical relationship between {RSSI}, {MCS}, and spatial streams,'' in \emph{2019 Wireless Days (WD)}, 2019, pp. 1--8.

\bibitem{PLmodel_mmWave}
G.~R. MacCartney, S.~Deng, and T.~S. Rappaport, ``Indoor office plan environment and layout-based {mmWave} path loss models for 28 {GHz} and 73 {GHz},'' in \emph{IEEE 83rd Vehicular Technology Conference (VTC Spring)}, 2016, pp. 1--6.

\bibitem{PL_mmWave}
A.~I. Sulyman \emph{et~al.}, ``Directional radio propagation path loss models for millimeter-wave wireless networks in the 28-, 60-, and 73-{GHz} bands,'' \emph{IEEE Transactions on Wireless Communications}, vol.~15, no.~10, pp. 6939--6947, 2016.

\bibitem{beamtrain}
C.~R. C.~M. Da~Silva \emph{et~al.}, ``Beamforming training for {IEEE} 802.11 ay millimeter wave systems,'' in \emph{Information Theory and Applications Workshop (ITA)}, 2018, pp. 1--9.

\bibitem{surveybeam}
\BIBentryALTinterwordspacing
A.~M. Nor, S.~Halunga, and O.~Fratu, ``Survey on positioning information assisted {mmWave} beamforming training,'' \emph{Ad Hoc Networks}, vol. 135, p. 102947, 2022. [Online]. Available: \url{https://www.sciencedirect.com/science/article/pii/S1570870522001287}
\BIBentrySTDinterwordspacing

\bibitem{alejandro}
\BIBentryALTinterwordspacing
A.~Blanco, P.~J. Mateo, F.~Gringoli, and J.~Widmer, ``Augmenting {mmWave} localization accuracy through sub-6 {GHz} on off-the-shelf devices,'' in \emph{Proceedings of the 20th Annual International Conference on Mobile Systems, Applications and Services}, ser. MobiSys '22.\hskip 1em plus 0.5em minus 0.4em\relax New York, NY, USA: Association for Computing Machinery, 2022, p. 477–490. [Online]. Available: \url{https://doi.org/10.1145/3498361.3538920}
\BIBentrySTDinterwordspacing

\bibitem{smart}
\BIBentryALTinterwordspacing
S.~Shamoon \emph{et~al.}, ``Integrated sub-6 {GHz} and millimeter wave band antenna array modules for {5G} smartphone applications,'' \emph{AEU - International Journal of Electronics and Communications}, vol. 161, p. 154542, 2023. [Online]. Available: \url{https://www.sciencedirect.com/science/article/pii/S143484112300016X}
\BIBentrySTDinterwordspacing

\bibitem{ris}
\BIBentryALTinterwordspacing
J.~Rao \emph{et~al.}, ``A shared-aperture dual-band sub-6 {GHz} and {mmWave} reconfigurable intelligent surface with independent operation,'' \emph{arxiv preprint}, 2024. [Online]. Available: \url{https://arxiv.org/abs/2406.02975}
\BIBentrySTDinterwordspacing

\bibitem{SGD_survey}
H.~Wang \emph{et~al.}, ``A comprehensive survey on training acceleration for large machine learning models in {IoT},'' \emph{IEEE Internet of Things Journal}, vol.~9, no.~2, pp. 939--963, 2022.

\bibitem{Survey_DL}
\BIBentryALTinterwordspacing
J.~Verbraeken \emph{et~al.}, ``A survey on distributed machine learning,'' \emph{ACM Computing Surveys}, vol.~53, no.~2, Mar. 2020. [Online]. Available: \url{https://doi.org/10.1145/3377454}
\BIBentrySTDinterwordspacing

\bibitem{FL_survey}
S.~Abdulrahman \emph{et~al.}, ``A survey on federated learning: The journey from centralized to distributed on-site learning and beyond,'' \emph{IEEE Internet of Things Journal}, vol.~8, no.~7, pp. 5476--5497, 2021.

\bibitem{ADAM}
\BIBentryALTinterwordspacing
D.~P. Kingma and J.~Ba, ``Adam: A method for stochastic optimization,'' 2017. [Online]. Available: \url{https://arxiv.org/abs/1412.6980}
\BIBentrySTDinterwordspacing

\bibitem{az}
P.~Picazo-Martinez \emph{et~al.}, ``{IEEE} 802.11az indoor positioning with {mmWave},'' \emph{IEEE Communications Magazine}, pp. 1--7, 2023.

\bibitem{axcsi}
\BIBentryALTinterwordspacing
F.~Gringoli, M.~Cominelli, A.~Blanco, and J.~Widmer, ``{AX-CSI}: Enabling {CSI} extraction on commercial 802.11ax {Wi-Fi} platforms,'' in \emph{Proceedings of the 15th ACM Workshop on Wireless Network Testbeds, Experimental Evaluation \& CHaracterization}, ser. WiNTECH '21.\hskip 1em plus 0.5em minus 0.4em\relax New York, NY, USA: Association for Computing Machinery, 2021, p. 46–53. [Online]. Available: \url{https://doi.org/10.1145/3477086.3480833}
\BIBentrySTDinterwordspacing

\bibitem{waveslam}
P.~Picazo \emph{et~al.}, ``{waveSLAM}: Empowering accurate indoor mapping using off-the-shelf millimeter-wave self-sensing,'' in \emph{IEEE 98th Vehicular Technology Conference (VTC2023-Fall)}, 2023, pp. 1--7.

\bibitem{mdtrack}
\BIBentryALTinterwordspacing
Y.~Xie, J.~Xiong, M.~Li, and K.~Jamieson, ``md-track: Leveraging multi-dimensionality for passive indoor wi-fi tracking,'' in \emph{The 25th Annual International Conference on Mobile Computing and Networking}, ser. MobiCom '19.\hskip 1em plus 0.5em minus 0.4em\relax New York, NY, USA: Association for Computing Machinery, 2019. [Online]. Available: \url{https://doi.org/10.1145/3300061.3300133}
\BIBentrySTDinterwordspacing

\bibitem{liang2022creatingData}
W.~Liang \emph{et~al.}, ``Advances, challenges and opportunities in creating data for trustworthy {AI},'' \emph{Nature Machine Intelligence}, vol.~4, no.~8, pp. 669--677, 2022.

\bibitem{Huy2022BetterDataLabelling}
H.~Tu, Z.~Yu, and T.~Menzies, ``Better data labelling with {EMBLEM} (and how that impacts defect prediction),'' \emph{IEEE Transactions on Software Engineering}, vol.~48, no.~1, pp. 278--294, 2022.

\bibitem{Leon2018RenderGAN}
\BIBentryALTinterwordspacing
L.~Sixt, B.~Wild, and T.~Landgraf, ``{RenderGAN}: Generating realistic labeled data,'' \emph{Frontiers in Robotics and AI}, vol.~5, 2018. [Online]. Available: \url{https://www.frontiersin.org/journals/robotics-and-ai/articles/10.3389/frobt.2018.00066}
\BIBentrySTDinterwordspacing

\bibitem{yolox2021}
Z.~Ge \emph{et~al.}, ``{YOLOX}: Exceeding {YOLO} series in 2021,'' \emph{arXiv preprint arXiv:2107.08430}, 2021.

\bibitem{comparisonnew1}
J.~Choi, ``Sensor-aided learning for wi-fi positioning with beacon channel state information,'' \emph{IEEE Transactions on Wireless Communications}, vol.~21, no.~7, pp. 5251--5264, 2022.

\bibitem{comparisonnew2}
\BIBentryALTinterwordspacing
T.~Ropitault, A.~Sahoo, S.~Blandino, and N.~Golmie, ``\BIBforeignlanguage{en}{Overhead-free people counting in mmwave network using ieee 802.11bf passive sensing},'' \emph{\BIBforeignlanguage{en}{IEEE International Symposium on Personal, Indoor and Mobile Radio Communications, Kauna, HI, US}}, 2024-10-17 04:10:00 2024. [Online]. Available: \url{https://tsapps.nist.gov/publication/get_pdf.cfm?pub_id=957538}
\BIBentrySTDinterwordspacing

\bibitem{finger}
K.~Liu \emph{et~al.}, ``Towards robust {WiFi} fingerprint-based vehicle tracking in dynamic indoor parking environments: An online learning framework,'' \emph{IEEE Transactions on Mobile Computing}, vol.~22, no.~12, pp. 6970--6984, 2023.

\bibitem{finger1}
N.~Singh, S.~Choe, and R.~Punmiya, ``Machine learning based indoor localization using {Wi-Fi RSSI} fingerprints: An overview,'' \emph{IEEE Access}, vol.~9, pp. 127\,150--127\,174, 2021.

\bibitem{finger2}
M.~Anjum \emph{et~al.}, ``{RSSI} fingerprinting-based localization using machine learning in {LoRa} networks,'' \emph{IEEE Internet of Things Magazine}, vol.~3, no.~4, pp. 53--59, 2020.

\bibitem{surveyWiFi}
S.~Tan, Y.~Ren, J.~Yang, and Y.~Chen, ``Commodity {WiFi} sensing in ten years: Status, challenges, and opportunities,'' \emph{IEEE Internet of Things Journal}, vol.~9, no.~18, pp. 17\,832--17\,843, 2022.

\bibitem{WiFivision}
Y.~He, Y.~Chen, Y.~Hu, and B.~Zeng, ``{WiFi} {Vision}: Sensing, recognition, and detection with commodity {MIMO-OFDM} {WiFi},'' \emph{IEEE Internet of Things Journal}, vol.~7, no.~9, pp. 8296--8317, 2020.

\bibitem{efficientfi}
J.~Yang \emph{et~al.}, ``{EfficientFi}: Toward large-scale lightweight {WiFi} sensing via {CSI} compression,'' \emph{IEEE Internet of Things Journal}, vol.~9, no.~15, pp. 13\,086--13\,095, 2022.

\bibitem{multisense}
\BIBentryALTinterwordspacing
Y.~Zeng \emph{et~al.}, ``{MultiSense}: Enabling multi-person respiration sensing with commodity {WiFi},'' \emph{Proceedings of the ACM on Interactive, Mobile, Wearable and Ubiquitous Technologies}, vol.~4, no.~3, sep 2020. [Online]. Available: \url{https://doi.org/10.1145/3411816}
\BIBentrySTDinterwordspacing

\bibitem{ml1}
H.~Choi \emph{et~al.}, ``{Wi-CaL}: {WiFi} sensing and machine learning based device-free crowd counting and localization,'' \emph{IEEE Access}, vol.~10, pp. 24\,395--24\,410, 2022.

\bibitem{ml2}
I.~Ahmad, A.~Ullah, and W.~Choi, ``{WiFi}-based human sensing with deep learning: Recent advances, challenges, and opportunities,'' \emph{IEEE Open Journal of the Communications Society}, vol.~5, pp. 3595--3623, 2024.

\bibitem{mmwavesur}
B.~van Berlo, A.~Elkelany, T.~Ozcelebi, and N.~Meratnia, ``Millimeter wave sensing: A review of application pipelines and building blocks,'' \emph{IEEE Sensors Journal}, vol.~21, no.~9, pp. 10\,332--10\,368, 2021.

\bibitem{mmwavemobi}
\BIBentryALTinterwordspacing
T.~Gu \emph{et~al.}, ``{mmSense}: Multi-person detection and identification via {mmWave} sensing,'' in \emph{Proceedings of the 3rd ACM Workshop on Millimeter-Wave Networks and Sensing Systems}, ser. mmNets '19.\hskip 1em plus 0.5em minus 0.4em\relax New York, NY, USA: Association for Computing Machinery, 2019, p. 45–50. [Online]. Available: \url{https://doi.org/10.1145/3349624.3356765}
\BIBentrySTDinterwordspacing

\bibitem{rapid}
J.~Pegoraro \emph{et~al.}, ``{RAPID}: Retrofitting {IEEE} 802.11ay access points for indoor human detection and sensing,'' \emph{IEEE Transactions on Mobile Computing}, vol.~23, no.~5, pp. 4501--4519, 2024.

\bibitem{p2}
B.~Huang, G.~Mao, Y.~Qin, and Y.~Wei, ``Pedestrian flow estimation through passive {WiFi} sensing,'' \emph{IEEE Transactions on Mobile Computing}, vol.~20, no.~4, pp. 1529--1542, 2021.

\bibitem{p3}
W.~Li \emph{et~al.}, ``Passive {WiFi} radar for human sensing using a stand-alone access point,'' \emph{IEEE Transactions on Geoscience and Remote Sensing}, vol.~59, no.~3, pp. 1986--1998, 2021.

\bibitem{p4}
K.~Chetty, G.~E. Smith, and K.~Woodbridge, ``Through-the-wall sensing of personnel using passive bistatic {WiFi} radar at standoff distances,'' \emph{IEEE Transactions on Geoscience and Remote Sensing}, vol.~50, no.~4, pp. 1218--1226, 2012.

\bibitem{p5}
W.~Chang \emph{et~al.}, ``Online public transit ridership monitoring through passive {WiFi} sensing,'' \emph{IEEE Transactions on Intelligent Transportation Systems}, vol.~24, no.~7, pp. 7025--7034, 2023.

\bibitem{p6}
H.~Sun, L.~G. Chia, and S.~G. Razul, ``Through-wall human sensing with {WiFi} passive radar,'' \emph{IEEE Transactions on Aerospace and Electronic Systems}, vol.~57, no.~4, pp. 2135--2148, 2021.

\bibitem{p7}
Z.~Wang, J.~A. Zhang, M.~Xu, and Y.~J. Guo, ``Single-target real-time passive {WiFi} tracking,'' \emph{IEEE Transactions on Mobile Computing}, vol.~22, no.~6, pp. 3724--3742, 2023.

\end{thebibliography}

\end{document}